\documentclass[10pt, conference, letterpaper]{IEEEtran}
\IEEEoverridecommandlockouts
\usepackage{cite}
\usepackage{graphicx}
\usepackage{url}
\usepackage{amsmath}
\usepackage{latexsym,amssymb,epsfig,color}
\usepackage{algorithm}
\usepackage{algorithmic}
\usepackage{slashbox}
\usepackage[tight,footnotesize,center]{subfigure}
\usepackage{color}
\usepackage{multirow}

\newcommand{\BEQA}{\begin{eqnarray}}
\newcommand{\EEQA}{\end{eqnarray}}

\newcommand{\Pp}{\mathcal{P}}
\newcommand{\F}{\mathcal{F}}
\newcommand{\K}{\mathbb{K}}
\newcommand{\T}{\mathbb{T}}
\newcommand{\E}{\mathbb{E}}
\newcommand{\pr}{\mathbb{P}}
\newcommand{\R}{\mathbb{R}}
\newcommand{\HBT}{\hat{\boldsymbol{\theta}}}

\newcommand{\Ht}{\hat{\theta}}

\newtheorem{theorem}{Theorem}
\newtheorem{lemma}[theorem]{Lemma}
\newtheorem{corollary}[theorem]{Corollary}
\newtheorem{definition}{Definition}

\begin{document}
%
\title{Incentivizing Sharing in Realtime D2D Streaming Networks: A Mean Field Game Perspective}

\author{ Jian Li$^*,$ Rajarshi Bhattacharyya$^*,$ Suman Paul$^*,$ Srinivas Shakkottai$^*,$ and Vijay Subramanian$^{\dag}$\\ ${}^*$Dept. of ECE, Texas A\&M University, ${}^{\dag}$Dept. of EECS, University of Michigan\\ Email: ${}^*$\{lj0818, rajarshibh, sumanpaul, sshakkot\}@tamu.edu, ${}^{\dag}$vgsubram@umich.edu}


\maketitle
\begin{abstract} 
We consider the problem of streaming live content to a cluster of co-located wireless devices that have both an expensive unicast base-station-to-device (B2D) interface, as well as an inexpensive broadcast device-to-device (D2D) interface, which can be used simultaneously. Our setting is a streaming system that uses a block-by-block random linear coding approach to achieve a target percentage of on-time deliveries with minimal B2D usage.  Our goal is to design an incentive framework that would promote such cooperation across devices, while ensuring good quality of service.  Based on ideas drawn from truth-telling auctions, we design a mechanism that achieves this goal via appropriate transfers (monetary payments or rebates) in a setting with a large number of devices, and with peer arrivals and departures.  Here, we show that a Mean Field Game can be used to accurately approximate  our system.  Furthermore, the complexity of calculating the best responses under this regime is low.  We implement the proposed system on an Android testbed, and illustrate its efficient performance using real world experiments.
\end{abstract}

\section{Introduction}\label{sec:intro}
There has recently been much interest in networked systems for collaborative resource utilization.  These are systems in which agents contribute to the overall welfare through their individual actions.  Usually, each agent has a certain amount of resources, and can choose how much to contribute based on the perceived return via repeated interactions with the system. An example is a peer-to-peer file sharing network, wherein each peer can contribute upload bandwidth by transmitting chunks to a peer, and receive downloads of chunks from that peer as a reward.  Interactions are bilateral, and hence tit-for-tat type strategies are successful in preventing free-riding behavior \cite{Coh03}.  More generally, collaborative systems entail multilateral interactions in which the actions of each agent affect and are affected by the collective behavior of a subset of agents.  Here, more complex mechanisms are needed to accurately determine the value of the contribution of each individual to the group.

An example of a collaborative system with repeated multilateral interactions is a device-to-device (D2D) wireless network.  Suppose that multiple devices require the same content chunk.  The broadcast nature of the wireless medium implies that several agents can be simultaneously satisfied by a single transmission.  However, they might each have different values for that particular chunk, and may have contributed different amounts in the past to the transmitting agent.  Furthermore, D2D systems undergo ``churn'' in which devices join and leave different clusters as they move around.   How then is an agent to determine whether to collaborate with others, and whether it has received a fair compensation for its contribution?   

\begin{figure}[ht]
\begin {center}
\includegraphics[width=2.2in]{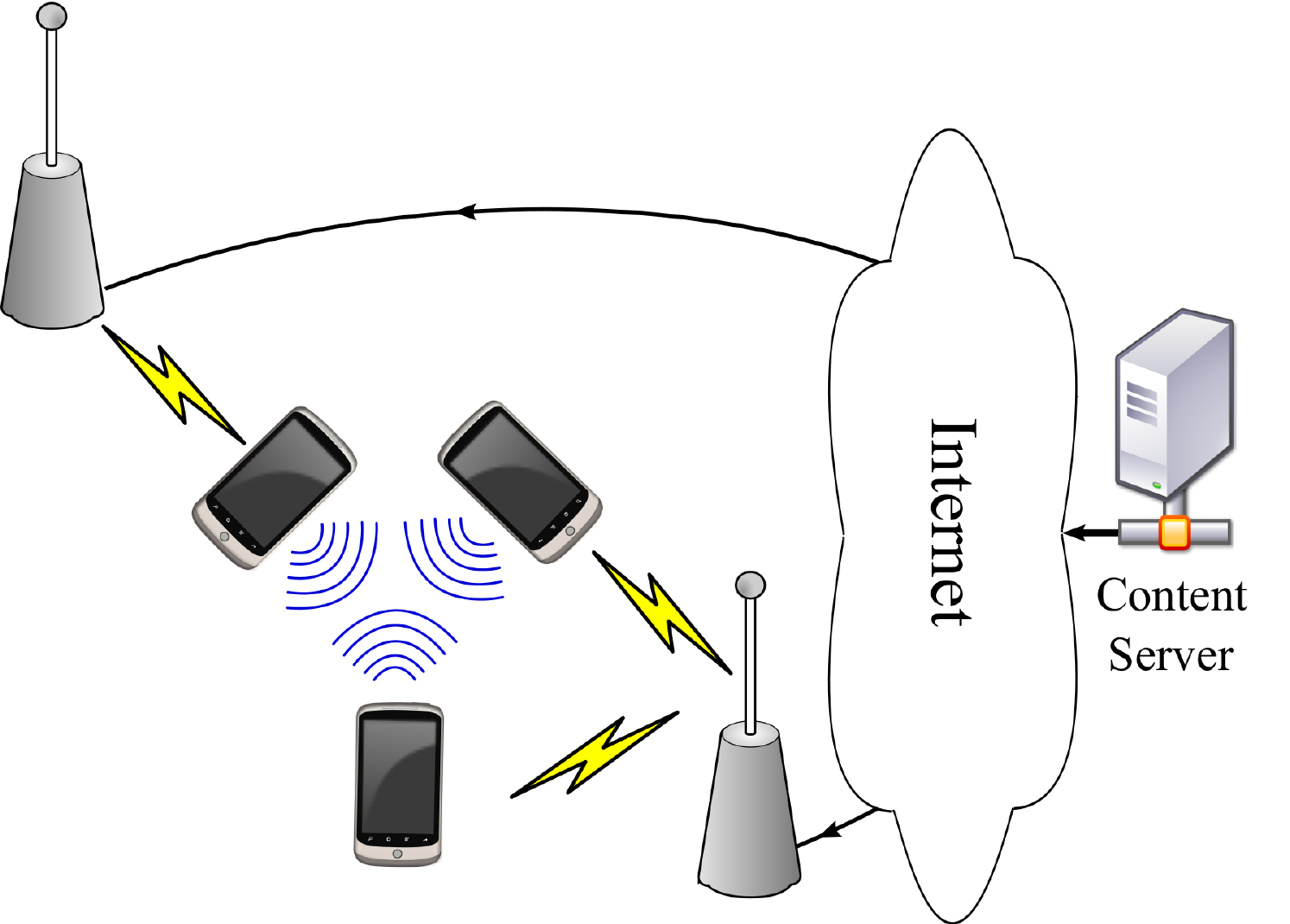}
\caption{Wireless content distribution via multiple interfaces  \cite{AbeSam13}.}
\label{fig:hybrid-p2p-android}
\end{center}
\end{figure}

Our objective in this paper is to design mechanisms for cooperation in systems with repeated multilateral interactions.  As in earlier literature, we assume that there exists a currency to transfer utility between agents \cite{gamenets,YuZho14}, and our goal is to determine how much should be transferred for optimal collaboration.  We focus on wireless content streaming as our motivating example.    In particular, as shown in Figure~\ref{fig:hybrid-p2p-android},  we assume that all devices are interested in the same content stream, and receive a portion of chunks corresponding to this stream via a unicast base-station-to-device (B2D) interface.  The B2D interface has a large energy and dollar cost for usage, and the devices seek to mitigate this cost via sharing chunks through broadcast D2D communication.\footnote{Note that, as we describe in greater detail later in this paper, it is possible to enable the usage of both the 3G (unicast) and WiFi (broadcast)  interfaces simultaneously on Android smart phones.}

\begin{figure}[ht]
\begin {center}
\includegraphics[width=2.8in]{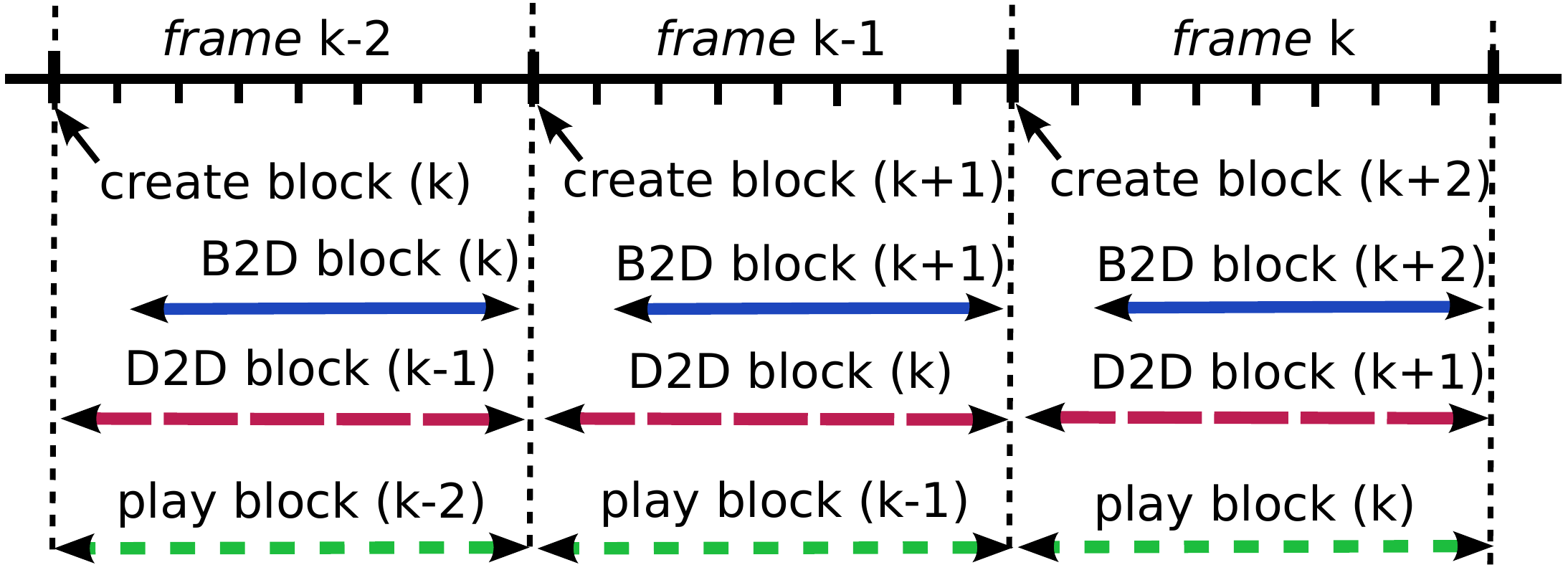}
\caption{Streaming architecture \cite{AbeSam13} in which each block must be delivered within two frames after its creation.}
\label{fig:stream_timing}
\vspace{-0.15in}
\end{center}
\end{figure}

A content sharing system is described in \cite{AbeSam13}, in which the objective is to achieve \emph{live streaming} of content synchronously to multiple co-located devices.  The system architecture of that work forms an ideal setting for studying mechanism design in which multilateral interactions occur.    The setup is illustrated in Figure~\ref{fig:stream_timing}.  Here, time is divided into \emph{frames}, which are subdivided into $T$ \emph{slots.}  A \emph{block} of data is generated by the content server in each frame, and the objective is to ensure that this block can be played out by all devices two frames after its generation, \emph{i.e.}, data block $k$ is generated in frame $k-2,$ and is to be played out in frame $k$.   Such a strict delay constraint between the time of generation and playout of each data block ensures that the \emph{live}  aspect of streaming is maintained. 

Upon generation of block $k$ (in frame $k-2$), the content server divides it into $N$ \emph{chunks} and performs random linear coding (RLC) over these chunks \cite{DebMed06}.  The server unicasts some of these coded chunks to each device using its B2D interface.  This number is to be kept small to reduce B2D usage.   Next, in frame $k-1,$ the devices use the broadcast D2D network to disseminate these chunks among themselves.   At the end of frame $k-1$ the devices attempt to decode block $k.$  If  a device $i$ has received enough coded chunks to decode the block, it plays out that block during frame $k.$  Otherwise, $i$ will be idle during this frame.   The use of RLC results in two desirable system features.  \emph{First}, the server can unicast a fixed number of chunks to the devices in each frame over a lossy channel (Internet plus B2D link) without any feedback.   \emph{Second}, the devices do not need to keep track of what chunks each one possesses while performing D2D broadcasts. 

The notion of quality of experience (QoE) here is \emph{delivery-ratio} denoted by $\eta$, which is the average ratio of blocks desired to the blocks generated \cite{HouBor09}.  For instance, a delivery ratio of $95\%$ would mean that it is acceptable if $5\%$ of the blocks can be skipped. A device can keep track of its QoE thus far  via the ``deficit'' incurred upto frame $k,$ which is the difference between the actual number of number of blocks successfully decoded by frame $k$ and the target value $\eta k.$ In \cite{AbeSam13}, it was shown that, assuming complete cooperation by the participating devices, it is possible to design a chunk sharing scheme whereby all devices would meet their QoE targets with minimal usage of the B2D interface.  \emph{But how do we design a mechanism to ensure that the devices cooperate?}

The setting of interest in this paper is that of  a large number of D2D clusters, each with a fixed number of agents, and with all clusters interested in the same content stream.  Examples of such settings are sports stadia, concerts or protest meetings, where a large number of agents gather together, and desire to receive the same live-stream (replays, commentary, live video \emph{etc.})  Devices move between clusters as agents move around, causing churn.    The objective of our work is to develop an incentive framework wherein each device truthfully reports the number of chunks that it receives via B2D and its deficit in each frame, so that a system-wide optimal allocation policy can be employed.  Such an incentive framework should be lightweight and compatible with minimal amounts of history retention. Finally, we also desire to implement the system on Android smart phones and measure its real world performance.

\subsection*{Related Work}

The question of how to assign value to wireless broadcast transmissions is intriguing.    For instance, \cite{HouLiu13} considers a problem of repeated interaction with time deadlines by which each node needs to receive a packet.  Each node declares its readiness to help others after waiting for a while; the game lies in choosing this time optimally, and the main result is to characterize the price of anarchy that results.  However, decision making is myopic, \emph{i.e.}, devices do not estimate future states while taking actions.  In a similar fashion, \cite{YuZho14} propose a scheme for sharing 3G services via WiFi hotspots using a heuristic scheme that possesses some attractive properties.  Here too, decision making is myopic. The question of fair scheduling at a base station that uses the history of interactions with individual stations in order to identify whether they are telling the truth about their state is considered in \cite{KavAlt14}.  However, since the devices in our network undergo churn and keeping track of device identities is infeasible, we desire a scheme that does not use identities or history to enable truthful revelation of state. The initial version of this work was presented in \cite{LiBha15} in which all proofs were omitted due to space constraints.  This paper presents complete details of the analytical methodology.

\subsubsection*{Perfect Bayesian and Mean Field Equilibria}\label{sec:PBE-MFE}

The typical solution concept in dynamic games is that of the Perfect Bayesian Equilibrium (PBE). Consider a strategy profile for all players, as well as beliefs about the other players' types at all information sets. This strategy profile and belief system is a PBE if: (i) \emph{Sequential rationality}: Each player's strategy specifies optimal actions, given her beliefs and the strategies of other players; (ii) \emph{Consistency of beliefs}: Each player's belief is consistent with the strategy profile (following Bayes' rule).
The PBE requires each agent to keep track of their beliefs on the future plays of all other agents in the system, and play the best response to that belief. The dynamic pivot mechanism \cite{BerVal10} extends the truth-telling VCG idea \cite{Kri02} to dynamic games.  It provides a basis for designing allocation schemes that are underpinned by truthful reporting. Translating the model in \cite{KavAlt14} to the language of \cite{BerVal10}, it is possible to use the dynamic pivot mechanism to develop a scheme (say FiniteDPM) with appropriate transfers that will be efficient, dominant strategy incentive compatible and per-period individually rational; note that while this scheme would use the identities of the devices, it will not need to build up a history of interactions. We omit the details of this as it is a straight-forward application of the general theory from \cite{BerVal10}.

Computation of PBE becomes intractable when the number of agents is large. An accurate approximation of the Bayesian game in this regime is that of a Mean Field Game (MFG) \cite{LasLio07,Boyan88,huang2006large}.  In MFG, the agents assume that each opponent would play an action drawn \emph{independently} from a static distribution over its action space.  The agent chooses an action that is the best response against actions drawn in this fashion.  The system is said to be at Mean Field Equilibrium (MFE) if this best response action is itself a sample drawn from the assumed distribution, \emph{i.e.}, the assumed distribution and the best response action are consistent with each other \cite{IyeJoh14,ManRam14,JianBai15}.   Essentially, this is the canonical problem in game theory of showing the existence of a Nash equilibrium, as it applies to the regime with a large number of agents. We will use this concept in our setting where there are a large number of peer devices with peer churn.  

To the best of our knowledge, there is no prior work that considers mechanism design for multilateral repeated games in the mean field setting. One of the important contributions of this paper is in providing a truth-telling mechanism for a mean-field game. In the process of developing the mechanism we will also highlight the nuances to be considered in the mean-field setting. In particular, we will see that aligning two concepts of value---from the system perspective and from that of the agents---is crucial to our goal of truth-telling.

\subsection*{Organization and Main Results}

We describe our system model in Section \ref{sec:stream-model}.  Our system consists of a large number of clusters, with agents moving between clusters.  The lifetime of an agent is geometric; an agent is replaced with a new one when it exits.  Each agent receives a random number of B2D chunks by the beginning of each frame, which it then shares using D2D transmissions.  

In Section \ref{sec:system-model}, we present an MFG approximation of the system, which is accurate when the number of clusters is large.   Here, the agents assume that the B2D chunks received and deficits of the other agents would be drawn independently from some distributions in the future, and optimize against that assumption when declaring their states.  The objective is to incentivize agents to truthfully report their states (B2D chunks and deficit) such that a schedule of transmissions (called an ``allocation'') that minimizes the discounted sum of costs can be used in each frame.   The mechanism takes the form of a scheme in which tokens are used to transfer utility between agents.    A nuance of this regime is that while the system designer sees each cluster as having a new set of users (with IID states) in each time frame, each user sees states of all its competitors \emph{but not itself} as satisfying the mean field distribution.  Reconciling the two view points is needed to construct a cost minimizing pivot mechanism, whose truth-telling nature is shown in Section~\ref{section2}.  This is our main contribution in this paper.  The allocation itself turns out to be computationally simple, and follows a version of a min-deficit first policy \cite{AbeSam13}.

Next, in Sections \ref{sec:mfe-1}--\ref{sec:mfe-2}, we present details on how to prove the existence of the MFE  in our setting.  
Although this proof is quite involved, it follows in a high-level sense in the manner of \cite{IyeJoh14,ManRam14}.  We then turn to computing the MFE and the value functions needed to determine the transfers in Section \ref{sec:simulations}.  The value iteration needed to choose allocation is straightforward.

We present details of our Android implementation of a music streaming app used to collect real world traces in Section \ref{sec:android}.  
We discuss the viability of our system in Section \ref{sec:viability}, and illustrate that under the current price of cellular data access, our system provides sufficient incentives to participate.  Finally, we conclude in Section \ref{sec:conclusion}.

\section{Content Streaming Model}\label{sec:stream-model}

We consider a large number of D2D clusters, each with a fixed number of agents, and with all clusters interested in the same content stream.  We assume that a cluster consists of $M$ co-located peer devices denoted by  $i\in\{1,\ldots,M\}$\footnote{Our analysis is essentially unchanged when there are a random but finite number of devices in each cluster.}.  The data source generates the stream in the form of a sequence of blocks. Each block is further divided into $N$ chunks for transmission.  We use random linear network coding over the chunks of each block (with coefficients in finite field $F_q$ of size $q$).  We assume that the field size is very large; this assumption can be relaxed without changing our cooperation results.  Time is divided into frames, which are further divided into slots.   At each time slot $\tau$, each device can simultaneously receive up to one chunk on each interface.

\textit{B2D Interface:} Each device has a (lossy) B2D unicast channel to a base-station.  For each device $i$, we model the number of chunks received using the B2D interface in the previous frame by a random variable with (cumulative) distribution $\zeta$, independent of the other devices. The support of $\zeta$ is the set $\{0, 1, \cdots, T\}$, denoted by $\mathbb{T}$.  The statistics of this distribution depend on the number of chunks transmitted by the server and the loss probability of the channel.  In \cite{AbeSam13}, a method for calculating statistics based on the desired quality of service is presented.  We take the distribution $\zeta$ as given.

\textit{D2D Interface:} Each device has a zero-cost D2D broadcast interface, and only one device can broadcast over the D2D network at each time $\tau$.  For simplicity of exposition, we will assume that the D2D broadcasts are always successful; the more complex algorithm proposed in \cite{AbeSam13} to account for unreliable D2D is fully consistent\footnote{We will discuss this at the end of Section \ref{sec:mfe-proof}.} with our incentive scheme.  Since each D2D broadcast is received by all devices, there is no need to rebroadcast any information.  It is then straightforward to verify that the order of D2D transmissions does not impact performance.  Thus, we only need to keep track of the number of chunks transmitted over the D2D interfaces during a frame in order to determine the final state of the system. 

\textit{Allocation:} We denote the total number of coded chunks of block $k$ delivered to device $i$ via the B2D network during frame $k-2$ using $e_i[k] \sim \zeta.$    We call the vector consisting of the number of transmissions by each device via the D2D interfaces over frame $k-1$ as the ``allocation'' pertaining to block $k,$ denoted by $\mathbf{a}[k].$   Also, we denote the number received chunks of block $k$ by device $i$ via D2D during frame $k-1$  using $g_i[k].$  Due to the large field size assumption, if $e_i[k] + g_i[k] = N,$ it means that block $k$ can be decoded, and hence can be played out.  For simplicity of exposition, we develop our results assuming that the allocation is computed in a centralized fashion in each cluster.  However, we actually implement a distributed\footnote{At the end of Section \ref{sec:mfe-proof} we will argue that the distributed implementation is also consistent with our incentive scheme.} version on the testbed. 

\textit{Quality of Experience:} Each device $i$ has a delivery ratio $\eta_i\in (0,1],$ which is the minimum acceptable long-run average number of frames device $i$ must playout.  
In the mobile agents model, we assume that all devices have the same delivery ratio $\eta$ for simplicity.  
It is straightforward to extend our results to the case where delivery ratios are drawn from some finite set of values.  
The device keeps track of the current deficit using a deficit queue with length $d_i[k] \in \K.$  The set of possible deficit values is given by $\K=\big\{ k\eta-m: k, m \geq 0,  m\leq \lfloor k \eta \rfloor \big\},$ 
where for $x \in \R$, $\lfloor x \rfloor=\max\{ k\in \mathbb{Z}: k \leq x \}$ is the largest whole number that $x$ is greater than. Note that $\K$ is a countable set and the possible 
deficit values are all non-negative. In fact, by the well-ordering principle $\K$ can be rewritten as $\{d_n\}_{n\in\mathbb{N}}$ with $d_n$ an increasing sequence (without bound) such that $d_1=0$. 
We will use this representation to enumerate the elements of $\K$.
If a device fails to decode a particular block, its deficit increases by $\eta,$ else it decreases by $1-\eta.$  The impact of deficit on the user's quality of experience is modeled by a function $c(d_i[k]),$ which is convex, differentiable and monotone increasing.  The idea is that user unhappiness increases more with each additional skipped block.  

\textit{Transfers:} We asume the existence of a currency (either internal or a monetary value) that can be used to transfer utility between agents \cite{gamenets,YuZho14}. In our system, a negative transfer is a price paid by the agent, while a positive value indicates that the agent is paid by the system. Such transfer systems are well established; see for instance a review in \cite{gamenets}.  Transfers are used by agents either to pay for value received through others' transmissions, or to be compensated for value added to others by transmitting a chunk.  We assume that the transmissions in the system are monitored by a reliable device, which can then report these values to decide on the transfers.  In practice we use the device that creates each ad-hoc network as the monitor.  

An \emph{allocation policy} maps the  values of the B2D chunks received and deficits as revealed by agents, denoted by $\hat{\boldsymbol{\theta}}[k]:=(\hat{\mathbf{e}}[k],\hat{\mathbf{d}}[k-1]),$ to an allocation for that frame $\mathbf{a}[k].$  Given an allocation, agents have no incentive to deviate from it, since an agent that does not transmit the allocated number of chunks would see no benefit;  those time slots would have no transmissions by other agents either.  The fundamental question is that of how to incentivize the agents to reveal their states truthfully so that the constructed allocation can maximize system-wide welfare.

\section{Mean Field Model and Mechanism Design}\label{sec:system-model}
Our system consists of $JM$ agents (or users) organized into $J$ clusters with $M$ agents per cluster. As mentioned earlier, time is slotted into frames. At the end of a frame, any agent $i$ can leave the system only to be replaced by  a new agent (also denoted by $i$) whose initial deficit is drawn from a (cumulative) distribution $\Psi$ with support
$\K$. This event occurs with probability $\bar{\delta}=(1-\delta)$ independently for each agent, so that the lifetimes of the agents are geometrically distributed. As described in the previous section, we assume that the number of chunks received via B2D for agent $i$ in frame $k,$ denoted by $e_i[k],$ is chosen in an \emph{i.i.d.} fashion according to the (cumulative) distribution $\zeta,$ with support $\T$; one such distribution is the binomial distribution.  In addition to the agents having geometrically distributed lifetimes, we also allow mobility in our set-up. In particular, in every frame we assume that all the agents are randomly permuted and then assigned to clusters such that there are exactly $M$ agents in each cluster. Using this system as a starting point we will develop our mean-field model that will be applicable when the number of clusters $J$ is extremely large.   

\begin{figure}[ht]
\begin {center}
\vspace{-0.05in}
\includegraphics[width=3.4in]{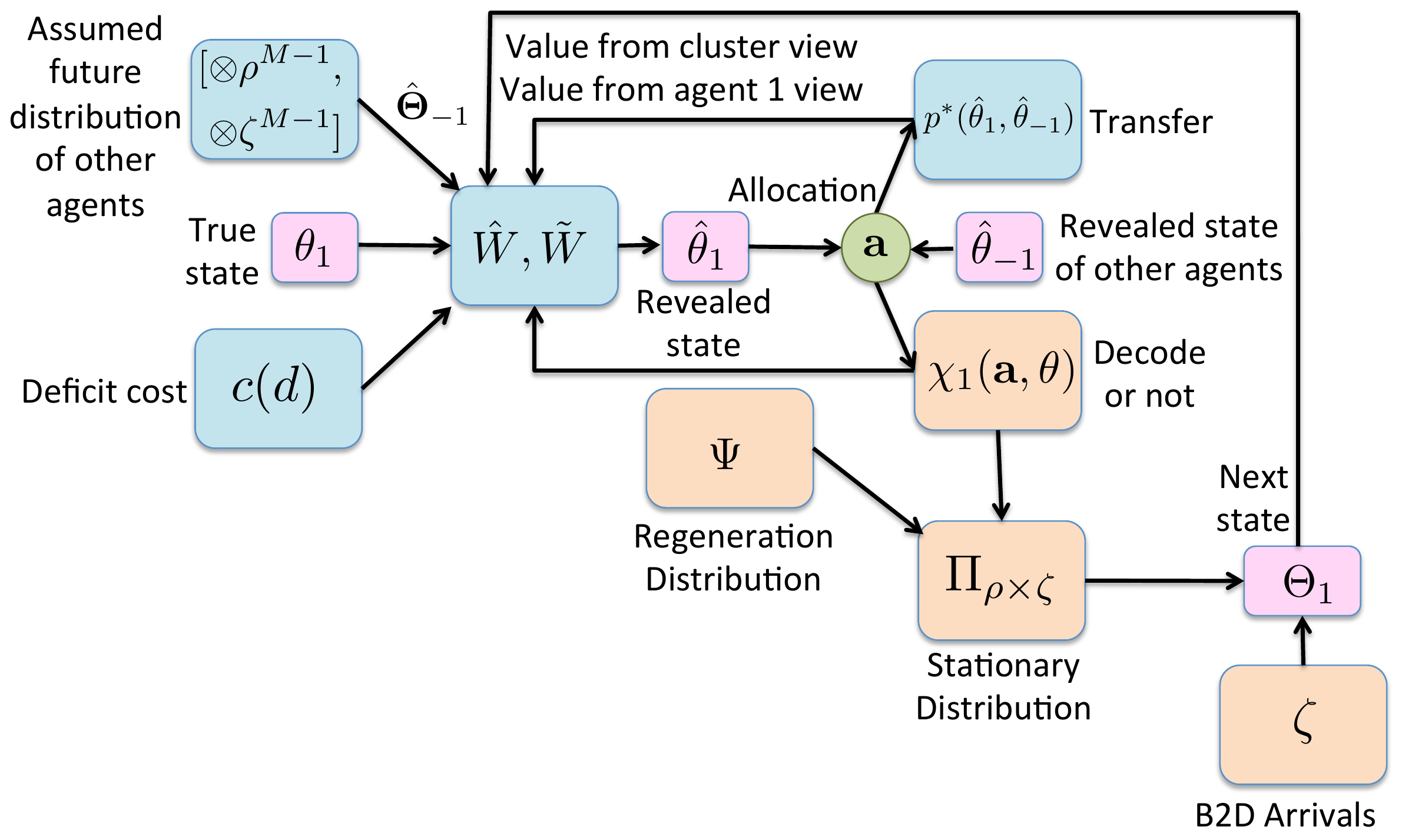}
\vspace{-0.1in}
\caption{The Mean Field system from perspective of agent $1.$}
\label{fig:MFE2}
\vspace{-0.15in}
\end{center}
\end{figure}

The mean field framework in Figure~\ref{fig:MFE2} illustrates system relationships that will be discussed below.  The blue/dark tiles apply to the value determination process for mechanism design, which will be discussed in this section.  The  beige/light tiles are relevant to showing the existence of an MFE on which the mechanism depends, which will be discussed in Sections~\ref{sec:mfe-1}--\ref{sec:mfe-2}.   

The mean field model yields  informational and computational savings, since otherwise each agent will need to not only be cognizant of the values and actions of all agents, but also track their mobility patterns.  Additionally, the mean field distribution accounts for regenerations, which do not have to be explicitly accounted for when determining best responses.

There is, however, an important nuance that the mean-field analysis introduces: when there are a large number of clusters, each cluster sees a different group of agents in every frame with their states drawn from the mean-field distribution, but even though each agent interacts with a new set of agents in every frame, it's own state is updated based on the allocations made to it, so that the differing viewpoints of the two entities need to be reconciled while providing any incentives.

The number of chunks received over the B2D interface and the deficit value constitute the state of an agent at the beginning of a frame. At frame $k$ we collect together the state variables of all the agents in system as $\boldsymbol{\theta}[k]=(\mathbf{e}[k], \mathbf{d}[k-1])$. Our mechanism then aims to achieve 
\begin{equation}\small 
W(\boldsymbol{\theta}[k])=\min_{\{\mathbf{a}[l]\}_{l=k}^{\infty}} \E\left\{ \sum_{j=1}^J \sum_{l=k}^{\infty} {\delta}^{l-k} \sum_{i\in s_j[l]} v_i(\mathbf{a}_{s_j}[l],{\theta}_i[l]) \right\},
\label{eq:mobile_all}
\end{equation}
where $j=1, 2, \cdots, J$ is the number of clusters in the system, $s_j[k]$ is the set of agents in cluster $j$ at frame $k$, $\mathbf{a}_{s_j}$ is the allocation in cluster $j$ and $v_i(\mathbf{a}_{s_j}[l],{\theta}_i[l])$ is the value that agent $i$ makes from the allocation in frame $k$. For agent $i$ set $j_i[k]$ to be the cluster he belongs in during frame $k$, i.e., $i\in s_{j_i[k]}[k]$. Note that the probability of remaining in the system $\delta$ appears as the discount factor in the above expression.

Given the allocation in each cluster, if agent $i$ does not regenerate, then his deficit gets updated as
\begin{equation}
d_i[k]= (d_i[k-1]+\eta - \chi_i(\mathbf{a}_{j_i[k]}[k],\theta_i[k]))^+,
\label{equ:dificit}
\end{equation}
where $(\cdot)^+=\max(\cdot,0)$, whereas if the agent regenerates, then $d_i[k]=\tilde{d}_i[k]$ where $\tilde{d}_i[k]$ is drawn \emph{i.i.d.} with distribution $\Psi$. Here, 
\begin{align}
& \chi_i(\mathbf{a},{\theta}_i)=1_{\{e_i+g_i(\mathbf{a}) = N\}} = \begin{cases}
1 & \text{if } e_i+g_i(\mathbf{a}) = N \\
0 & \text{otherwise,}
\end{cases}
\end{align}
where $\chi_i(.)$ is $1$ if and only if agent $i$ obtains all $N$ coded chunks to be able to decode a block, $g_i(\mathbf{a})$ is the number of packets agent $i$ can get during a frame under the allocation $\mathbf{a}$ (where we suppress the dependence of $\mathbf{a}$ on $\boldsymbol{\theta}$). We specialize to the case where the value per frame for agent $i$ with system state $\boldsymbol{\theta}$ and vector of allocations $\mathbf{a}$ is given by $v_i(\mathbf{a},\theta_i)=c\Big(\big(d_i+\eta - \chi_i(\mathbf{a},{\theta}_i)\big)^+\Big)$ if there is no regeneration and $v_i(\mathbf{a},\boldsymbol{\theta}_i)=c(\tilde{d}_i)$ otherwise, where $\tilde{d}_i$ is \emph{i.i.d.} with distribution $\Psi$ and $c(\cdot)$ is the holding cost function that is assumed to be convex and monotone increasing.

As there are a large number of clusters, in every frame there is a completely different set of agents that appear at any given cluster.  The revealed states of these agents will be drawn from the mean field distribution.  Hence, from the perspective of some cluster $l,$ the revealed state of the agents in that cluster $\hat{\boldsymbol{\Theta}}_{l}$ will be drawn 
according to the (cumulative) distributions $[\otimes \rho^M, \otimes \zeta^M],$ with $\rho$ pertaining to the deficit, and $\zeta$ pertaining to the B2D transmissions received by that agent. 
 Note that the support of $\rho$ is $\K$ while the support of $\zeta$ is $\T,$ and $\otimes$ indicates the \emph{i.i.d} nature of the agent states.  Whereas from the perspective a particular agent $i,$ the revealed states of all the other agents in that cluster will be drawn according to $\hat{\boldsymbol{\Theta}}_{-i} \sim [\otimes \rho^{M-1}, \otimes \zeta^{M-1}].$  These facts will simplify the allocation problem in each cluster and also allow us to analyze the MFE by tracking a particular agent. 

First, we consider the allocation problem as seen by the clusters.  Pick any finite number of clusters. In the mean-field limit, the agents from frame to frame will be different in each cluster, therefore the allocation decision in each cluster can be made in an distributed manner, independent of the other clusters; this is one of the chaos hypotheses of the mean-field model. This then implies that the objective in \eqref{eq:mobile_all} is achieved by individual optimization in each cluster, \emph{i.e.}, 
\begin{align}
W(\HBT [k])=\sum_{j=1}^J W_j(\HBT_{s_j}[k]),
\end{align}
where we recall that $\HBT_{s_j}[k]$ is the revealed state of agents in cluster $j$ at time $k$ and 
\begin{align}
W_j(\HBT_{s_j}[k])=\min_{\{\mathbf{a}_{s_j}[l]\}_{l=k}^{\infty}} \sum_{l=k}^{\infty} {\delta}^{l-k} \sum_{i\in s_j[l]} v_i(\mathbf{a}_{s_j}[l],{\Ht}_i[l]).
\end{align}
Under mean field assumption, the method of determining value does not change from step-to-step. The value function in the mean-field is determined by the first solving the following Bellman equation
\begin{align}
\hat{W}(\hat{\boldsymbol{\theta}})=\min_{\mathbf{a}} \sum_{i=1}^M v_i(\mathbf{a},{\Ht}_i) + \delta \E\left\{ \hat{W}(\hat{\boldsymbol{\Theta}})\right\}
\label{eq:valueMF}
\end{align}
to obtain function $\hat{W}(\cdot)$, where ${\hat{\boldsymbol{\theta}}}$ is the $M$-dimensional revealed state vector (with elements $\Ht_i$) and the future revealed state vector $\hat{\boldsymbol{\Theta}}$ is chosen according to $[\otimes \rho^{M}, \otimes \zeta^{M}]$, and thereafter setting $W_j(\HBT_{s_j}[k])=\hat{W}(\HBT_{s_j}[k])$ for every $j=1,2,\dotsc,J$. This observation then considerably simplifies the allocation in each cluster to be the greedy optimal, \emph{i.e.}, determine (multi)function
\begin{align}
\mathbf{a}^*(\hat{\boldsymbol{\theta}})=\arg\min_{\mathbf{a}} \sum_{i=1}^M v_i(\mathbf{a},{\Ht}_i),
\label{eq:allocMF}
\end{align}
and for $j=1,2,\dotsc,J$ we set $\mathbf{a}^*_{s_j}=\mathbf{a}^*(\hat{\boldsymbol{\theta}}_j)$.

Next, we consider the system from the viewpoint of a typical agent $i$; w.l.o.g let $i=1$.  Any allocation results in the deficit changing according to \eqref{equ:dificit} and the future B2D packets drawn according to $\zeta,$ whereas the state of every other agent that agent $1$ interacts with in the future gets chosen according to the mean field distribution.   Then the value function (of the cluster) from the perspective of agent $1$ is determined using \begin{align}
\begin{split}
& \tilde{W}(1,(\hat{\theta}_1,\hat{\boldsymbol{\theta}}_{-1})) = \\
& \min_{\mathbf{a}} \sum_{i^\prime=1}^M v_{i^\prime}(\mathbf{a},\Ht_{i^\prime}) + \delta \E\left\{\tilde{W}(1,(\hat{\Theta}_1, \hat{\boldsymbol{\Theta}}_{-1})) | \mathbf{a}, \hat{\theta}_1\right\}.
\end{split}
\label{eq:valueAgentMF}
\end{align} 
Here, $\hat{\boldsymbol{\theta}}_{-1}$ represents the revealed states of all the agents in cluster except $1,$   $\hat{\boldsymbol{\Theta}}_{-1} \sim [\otimes \rho^{M-1}, \otimes \zeta^{M-1}],$ and for $\hat{\Theta}_1,$ the deficit term is determined via \eqref{equ:dificit} (setting $\theta_i = \hat{\theta}_i$) while the B2D term follows $\zeta.$  This recursion yields a function $\tilde{W}(1,\cdot)$ which applies to all agents. Using this function, one can also determine the allocation that agent $1$ expects his cluster to perform, namely, 
\begin{align}
\begin{split}
& \tilde{\mathbf{a}}(\Ht_1,\hat{\boldsymbol{\theta}}_{-1})=\\
& \arg\min_{\mathbf{a}} \sum_{i^\prime=1}^M v_{i^\prime}(\mathbf{a},\Ht_{i^\prime}) + \delta \E\left\{\tilde{W}(1,(\hat{\Theta}_1, \hat{\boldsymbol{\Theta}}_{-1})) | \mathbf{a}, \Ht_1\right\}.
\end{split}
\label{eq:allocAgentMF}
\end{align} 

Using the two allocations $\mathbf{a}^*$ and $\tilde{\mathbf{a}}$ we can write down the value of agent $1$ from the system optimal allocation and the value of agent $1$ in the allocation that the agent thinks that the system will be performing. For a given allocation function $\mathbf{a}(\cdot)$ (for the state of agents in the cluster where agent $1$ resides at present), we determine the solution to the following recursion
\begin{align}\label{eqn:true-val}
V(\mathbf{a}(\hat{\boldsymbol{\theta}}),\tilde{\theta}_1)=v_1(\mathbf{a},\tilde{\theta}_1) + \delta \E\left\{V(\mathbf{a}(\hat{\Theta}_1,\hat{\boldsymbol{\Theta}}_{-1}),\tilde{\Theta}_1)\right\}
\end{align}
to get function $V(\cdot,\cdot)$,  where $\tilde{\theta_1},$ is an arbitrary state variable, the deficit term of $\tilde{\Theta}_1$ follows \eqref{equ:dificit} while the B2D term is generated independently (setting $\theta_i = \tilde{\theta}_i$), $\mathbf{a}$ is an arbitrary allocation, the B2D term is generated independently, and $\hat{\boldsymbol{\Theta}}_{-1}$ is chosen using the mean-field distribution.  Notice that $\tilde{\theta}_1 = \theta_1$ would yield the true value of allocation $\mathbf{a}$  to agent $1.$  By the cluster optimal allocation (what the cluster actually does), agent $1$ gets $V(\mathbf{a}^*(\Ht_1,\hat{\boldsymbol{\theta}}_{-1}),\theta_1)$ whereas from the perception of agent $1$ he thinks he should be getting $V(\tilde{\mathbf{a}}(\Ht_1,\hat{\boldsymbol{\theta}}_{-1}),\theta_1)$ (based on what he thinks the cluster should be doing).

\subsection*{Transfer}
We will use the different value functions to define the transfer for agent $1$ depending on the reported state variable $\hat{\theta_1}$ such that the  transfer depends on the difference between what he gets from the system optimal allocation and what he expects the system to do from his own perspective. Using this logic we set the transfer for agent $1$ as
\begin{align}
\begin{split}
p^*(\Ht_1,\hat{\boldsymbol{\theta}}_{-1}) =V(\mathbf{a}^*(\hat{\boldsymbol{\theta}}),\Ht_1) - V(\tilde{\mathbf{a}}(\hat{\boldsymbol{\theta}}),\Ht_1) \hspace{0.5in}\\
 +H(\hat{\boldsymbol{\theta}}_{-1}) - (\tilde{W}(1,(\hat{\theta}_1,\hat{\boldsymbol{\theta}}_{-1})) - V(\tilde{\mathbf{a}}(\hat{\boldsymbol{\theta}}),\Ht_1) ).   
 \end{split}
\label{eq:transferMF}
\end{align}
where $H(\hat{\boldsymbol{\theta}}_{-1}),$ following the Groves pivot mechanism, can be chosen using the recursion
\begin{align}
H(\hat{\boldsymbol{\theta}}_{-1})=\min_{\mathbf{a_{-1}}} \sum_{i\neq 1} v_i(\mathbf{a_{-1}},{\Ht}_i) + \delta \E\left\{ H(\hat{\boldsymbol{\Theta}}_{-1})\right\},
\label{eq:trans}
\end{align}
where $\hat{\boldsymbol{\Theta}}_{-1} \sim [\otimes \rho^{M-1}, \otimes \zeta^{M-1}],$ and $\mathbf{a_{-1}}$ is used to denote an allocation in a system in which agent $1$ is not present.

The Clarke pivot mechanism idea ensures that the net-cost of agent~$1$,  $V(\mathbf{a}^*(\hat{\boldsymbol{\theta}}),\Ht_1) - p^*(\Ht_1,\hat{\boldsymbol{\theta}}_{-1}),$ equals $\tilde{W}(1,(\hat{\theta}_1,\hat{\boldsymbol{\theta}}_{-1}))-H(\hat{\boldsymbol{\theta}}_{-1}).$  This is simply the value of the system as a whole from the viewpoint of agent~$1,$ minus a  function only of $\hat{\boldsymbol{\theta}}_{-1}.$  As in the Vickrey-Clarke-Groves mechanism, such formulation of net-cost naturally promotes truth-telling as a dominant strategy at each step.

\subsection*{Allocation Scheme}

The basic building block of our mechanism is the per-frame optimal allocations that solve \eqref{eq:mobile_all}. We will now spell out the allocation in greater detail. First, we observe that the allocation problem separates into independent allocation problems in each cluster that have the same basic structure. Therefore, it suffices to discuss the allocation problem for one cluster.

From \eqref{eq:allocMF}, the objective in this cluster is 
\begin{align}
\min_{\mathbf{a}} \sum_{i=1}^M c( (d_i[k-1]+\eta - \chi_i(\mathbf{a}[k],\theta_i[k]))^+) \label{eqn:opt-alloc}
\end{align}

An optimal allocation is determined using the following observations. First, we partition the agents into two sets, ones who cannot decode the frame even if they never transmit during the $T$ slots of the D2D phase and the rest; the former agents are made to transmit first. After this we determine agents who have extra chunks (number of slots that they can transmit on such that there is still time to decode whole frame) and make these agents transmit their extra chunks. After all the extra chunks have been transmitted, it is easy to see using the properties of the holding cost function that agents are made to transmit in a minimum-deficit-first fashion in order to prioritize agents with large deficits. This is summarized in the follow lemma.
\begin{lemma}\label{lem:opt_allocation}
The algorithm delineated in Algorithm~\ref{alg: D2D} provides an optimal greedy allocation.
\end{lemma}
\begin{IEEEproof}
Given the B2D arrivals $(e_1[k],...,e_M[k])$, we partition the set of devices $\{1,...,M\}$ into sets $\cal{S}$ and ${\cal{S}}^c=\{1,...,M\}\backslash {\cal{S}},$ based on whether $e_i[k]+T-N \geq 0$ or not.  Those agents that satisfy this condition can potentially receive enough chunks during the D2D phase that they can decode the block, whereas the others cannot.  Hence, all members of ${\cal{S}}^c$  can potentially transmit their chunks in the allocation solving \eqref{eqn:opt-alloc}.  Let $T_1=\min\{\sum_{i\in {\cal S}^c}e_i[k]$,\ T\}.  So we can devote the first $T_1$ slots of the current frame to transmissions from the devices in ${\cal S}^c$.

Let the number of transmissions made by agent $i$ in allocation $\mathbf{a}$  be denoted by $x_i[k].$  We can write down the constraints that any feasible allocation $\mathbf{a}$ must satisfy as
\BEQA
\begin{array}{ll}
0\leq x_i[k]\leq e_i[k] & \text{$\forall\ i\in {\cal S}$}\\
\sum_{i\in {\cal S}}x_i[k]= T-T_1 & 
\end{array}
\EEQA
Observe that each agent can transmit $e_i[k]+T-N$ chunks without affecting the above constraints (\emph{i.e.,} it does not change its chances of being able to decode the block, as there is enough time left for it to receive chunks that it requires).  We call these as ``extra'' chunks.  Suppose that all extra chunks have been transmitted by time $T_2<T,$ and no device has yet reached full rank.  At this point, all agents in the system need the same number of chunks, and any agent that transmits a chunk will not be able to receive enough chunks to decode the block.  In other words, agents now have to ``sacrifice'' themselves one at a time, and transmit all their chunks.  The question is, what is the order in which such sacrifices should take place? 

Compare two agents $i$ and $j,$ with deficits $d_i > d_j.$  Also, let $\chi \in\{0,1\}.$  Now, for either value of $\chi$ 
$$
d_i-(d_i - \chi)^+ \geq d_j-(d_j - \chi)^+.
$$
Hence, since $c(.)$ is convex and monotone increasing,
\begin{align}
&\int_{(d_i - \chi)^+}^{d_i} c'(z) dz \geq  \int_{(d_j - \chi)^+}^{d_j} c'(z) dz \geq 0 \\
\Rightarrow\quad  & c(d_i) - c((d_i - \chi)^+) \geq c(d_j) - c((d_j - \chi)^+) \geq 0.
\end{align}

Now, consider the following problem with $\chi_i,\chi_j \in \{0,1\}$ under the constraint $\chi_i + \chi_j =1:$ 
\begin{align}
& \min_{\chi_i,\chi_j} c(d_i -\chi_i) + c(d_j -\chi_j). \label{eqn:comparison}\\
\Leftrightarrow\quad  & \max_{\chi_i,\chi_j}  c(d_i) - c(d_i -\chi_i) + c(d_j) - c(d_j -\chi_j). 
\end{align}
Then, from the above discussion, the solution is to set $\chi_i =1$ and $\chi_j =0.$  Thus, comparing \eqref{eqn:comparison}  and \eqref{eqn:opt-alloc}, the final stage of the allocation should be for agents to sacrifice themselves according to a min-deficit-first type policy.  Algorithm~\ref{alg: D2D} describes the final allocation rule.
\end{IEEEproof}

\begin{algorithm}
\caption{Optimal Mean Field D2D Allocation Rule}
\label{alg: D2D}
\begin{algorithmic}
\STATE {\ At the beginning of each frame $k-1$, given the arrivals $(e_1[k],...,e_M[k])$:}
\STATE Partition the devices into sets ${\cal{S}}=\{i\in\{1,...,M\}:\ N-e_i[k]\leq T, e_i[k]+\sum_{j\neq i} e_j[k] \geq N\}$ and ${\cal S}^c$.
\STATE If ${\cal{S}}= \emptyset$, none of the agents can decode the block. Else,
\STATE {\bf Phase ${\mathbf 1)}$} Let all the agents in ${\cal S}^c$ transmit all that they initially received for the next $T_1=\min\{\sum\limits_{i\in{\cal S}^c}e_i[k], T\}$ slots.
\STATE If there exists time and a need for more transmissions, 
\STATE {\bf Phase ${\mathbf 2)}$} Let each agent $i\in {\cal S}$ transmit up to $(e_i[k]+T-N)^+$ of its initial chunks. 
\STATE {\bf Phase ${\mathbf 3)}$} While there exists time and a need for more transmissions,
let devices in ${\cal S}$ transmit their remaining chunks in an increasing order of their deficit values.
\end{algorithmic}
\end{algorithm}

\section{Properties of mechanism}\label{section2}
\subsection{Truth-telling as dominant strategy}
Since we consider a mean-field setting, we will assume that deficit of agent $i$ changes via the allocation while the deficits of all the other agents are drawn using the given distribution $\rho$. The $\mathbf{e}$ values are generated \emph{i.i.d.} with distribution $\zeta$.
Based on the system state report $\boldsymbol{\theta}[k]$ at time $k$,
we assume that the mechanism makes the optimal greedy allocation
$\mathbf{a}^*[k]$ from \eqref{eq:allocMF} and levies transfers
$\mathbf{p}^*[k]$ from \eqref{eq:transferMF} that uses the allocations
from the agent's perspective from \eqref{eq:allocAgentMF}.  We can
then show that truthfully revealing the state, \emph{i.e.,} $(d,e)$
values at the beginning of every frame is incentive compatible. 

\begin{definition}
A direct mechanism (or social choice function) $f=(a, p)$ is \emph{dominant strategy incentive compatible} if $\theta_i$ is a dominant strategy at $\theta_i$ for each $i$ and $\theta_i\in\Theta_i$, where $a(\cdot)$ is a decision rule and $p(\cdot)$ is a transfer function.
\end{definition}

\begin{theorem}
Our mechanism $\{\mathbf{a}^{*}[k], \mathbf{p}^{*}[k]\}_{k=0}^{\infty}$ is dominant strategy incentive compatible. 
\end{theorem}
\begin{IEEEproof}
The net-cost in frame $k$ for agent $i$ when reporting $\theta_i[k]$ versus $r_i[k]$ is given by
\begin{align}
\begin{split}
& V(\mathbf{a}^*(\hat{\boldsymbol{\theta}}_{s_{j_i[k]}}[k]),\theta_i[k])-p^*(\theta_i[k],\hat{\boldsymbol{\theta}}_{-i}[k]) = \\
&\tilde{W}(i,(\hat{\theta}_i[k],\hat{\boldsymbol{\theta}}_{-i}[k])) -H(\hat{\boldsymbol{\theta}}_{-i}[k]) \\
& \leq \tilde{W}(i,(r_i[k],\hat{\boldsymbol{\theta}}_{-i}[k])) -H(\hat{\boldsymbol{\theta}}_{-i}[k])\\
& = V(\mathbf{a}^*((r_i[k],\hat{\boldsymbol{\theta}}_{-i}[k])),\theta_i[k])-p^*(r_i[k],\hat{\boldsymbol{\theta}}_{-i}[k]),
\end{split}
\end{align}
where $\theta_i$ is the true type and $r_i$ is an arbitrary type; the equalities hold true due to the definition of value function and transfer; the last inequality follows by the optimality of allocation $\tilde{\mathbf{a}}(\theta_i,\hat{\boldsymbol{\theta}}_{-i})$ in cluster $s_{j_i[k]}$ maximizes the system utility from the perspective of agent $i$. Therefore, in every frame it is best for agent $i$ to report truthfully and this holds irrespective of the reports of the other agents.
\end{IEEEproof}
\subsection{Nature of transfers}
We now determine the nature of the transfers that are required to promote truth-telling. We will show that the transfers constructed in \eqref{eq:transferMF} are always non-negative, \emph{i.e.,} the system needs to pay the agents in order to participate.  In other words, each agent needs a subsidy to use the system, since it could simply choose not to participate otherwise.  Thus, the system is not budget-balanced. We will show later how the savings in B2D usage that results from our system provides the necessary subsidy in Section~\ref{sec:viability}.  Given these transfers, we will also see that our mechanism is individually rational so that users participate in each frame. 

\begin{lemma}\label{lem:posTx}
 The transfers defined in \eqref{eq:transferMF} are always non-negative. 
\end{lemma}
\begin{IEEEproof}
From \eqref{eq:transferMF}, we have 
\begin{align}
\label{eq:positivetransfer}
&p^*(\Ht_i[k],\hat{\boldsymbol{\theta}}_{-i}[k]) =V(\mathbf{a}^*(\hat{\boldsymbol{\theta}}[k]),\Ht_i[k])  \\
 &\quad +H(\hat{\boldsymbol{\theta}}_{-i}[k]) - \tilde{W}(i,(\hat{\theta}_i[k],\hat{\boldsymbol{\theta}}_{-i}[k]))\notag \\
&=V(\mathbf{a}^*(\hat{\boldsymbol{\theta}}[k]),\Ht_i[k])-V(\mathbf{a}_{-i}(\hat{\boldsymbol{\theta}}[k]_{-i}),\Ht_i[k]) \notag \\ 
&\; +\tilde{W}_{-i}(i,(\hat{\theta}_i[k],\hat{\boldsymbol{\theta}}_{-i}[k]))-\tilde{W}(i,(\hat{\theta}_i[k],\hat{\boldsymbol{\theta}}_{-i}[k])) \notag\\
&\stackrel{(a)}{\geq} V(\mathbf{a}^*(\hat{\boldsymbol{\theta}}[k]),\Ht_i[k])-V(\mathbf{a}_{-i}(\hat{\boldsymbol{\theta}}_{-i}[k]),\Ht_i[k]) \notag
\\ &
\stackrel{(b)}{\geq} 0. \notag
\end{align}
where $(a)$ follows from the definition of allocation $\tilde{\mathbf{a}}$ and the inequality $(b)$ is true by the monotonicity argument below. 

We assume that under both systems (with the allocations $\mathbf{a}_{-i}$ and $\mathbf{a}^*$), the deficits are initialized with the same value. Also note that all the agents follow the same reporting strategy in frame $k$, and hence, $\chi(\mathbf{a}^*)$ and $\chi(\mathbf{a}_{-i})$ can be compared.  Under allocation $\mathbf{a}_{-i}$, agent $i$ never transmits and will pick up free chunks from other agents' transmissions. However, agent $i$ may have to transmit under allocation $\mathbf{a}^*$. Thus, we have
\begin{align}
\chi_i(\mathbf{a}^*(\hat{\boldsymbol{\theta}}[k]), \theta_i[k]) \leq
\chi_i(\mathbf{a}_{-i}(\hat{\boldsymbol{\theta}}_{-i}[k]), \theta_i[k]),
\end{align}
as $e_i[k]+g_i(\mathbf{a}^*(\hat{\boldsymbol{\theta}}[k])) \leq
e_i[k]+g_i(\mathbf{a}_{-i}(\hat{\boldsymbol{\theta}}_{-i}[k]))$ is true for every $k$.

Using this we can compare the two deficits by considering the same allocation policy. For $k\geq 0$, we have 
\begin{align}
d_i(\mathbf{a}^*(\hat{\boldsymbol{\theta}}[k]))& =\big(d_i(\mathbf{a}^*(\hat{\boldsymbol{\theta}}[k-1]))+\eta-\chi_i(\mathbf{a}^*(\hat{\boldsymbol{\theta}}[k]), \theta_i[k])\big)^+
\end{align}
\begin{align}
d_i(\mathbf{a}_{-i}(\hat{\boldsymbol{\theta}}_{-i}[k]))& =\big( d_i(\mathbf{a}_{-i}(\hat{\boldsymbol{\theta}}_{-i}[k-1]))+\eta\\
& \qquad -\chi_i(\mathbf{a}_{-i}(\hat{\boldsymbol{\theta}}_{-i}[k]), \theta_i[k])\big)^+ \notag
\end{align}
with $\chi_i(\mathbf{a}^*(\hat{\boldsymbol{\theta}}[k]), \theta_i[k])\leq
\chi_i(\mathbf{a}_{-i}(\hat{\boldsymbol{\theta}}_{-i}[k]), \theta_i[k])$ for all $k$, which implies that $d_i(\mathbf{a}^*(\hat{\boldsymbol{\theta}}[k])) \geq
d_i(\mathbf{a}_{-i}(\hat{\boldsymbol{\theta}}_{-i}[k]))$. Since the function $V(\cdot,\cdot)$ in \eqref{eqn:true-val} can be obtained by value iteration starting with $v(\cdot)$, then by the definition of value function $v(\cdot)$ and the monotonicity of holding cost function $c(\cdot)$ in $d$, we have $V(\cdot,\cdot)$ being an increasing function in $d$. Then it directly follows that
\begin{align} 
 V(\mathbf{a}^*(\hat{\boldsymbol{\theta}}[k]),\Ht_i[k]) \geq
 V(\mathbf{a}_{-i}(\hat{\boldsymbol{\theta}}_{-i}[k]),\Ht_i[k]),
\end{align}
which completes our proof.
\end{IEEEproof}

The proof of individual rationality follows along the same lines as Lemma~\ref{lem:posTx}.
\begin{lemma}\label{lem:IR}
Our mechanism $\{\mathbf{a}^{*}[k], \mathbf{p}^{*}[k]\}_{k=0}^{\infty}$ is individually rational, i.e., the voluntary participation constraint is satisfied. 
\end{lemma}
\begin{IEEEproof}
The net-cost in frame $k$ for agent $i$ is given by
\begin{align}
\begin{split}
& V(\mathbf{a}^*(\hat{\boldsymbol{\theta}}_{s_{j_i[k]}}[k]),\theta_i[k])-p^*(\theta_i[k],\hat{\boldsymbol{\theta}}_{-i}[k]) \\
&=V(\mathbf{a}_{-i}(\hat{\boldsymbol{\theta}}_{-i}[k]),\theta_i[k]) \\
&-[\tilde{W}_{-i}(i,(\hat{\theta}_i[k],\hat{\boldsymbol{\theta}}_{-i}[k]))-\tilde{W}(i,(\hat{\theta}_i[k],\hat{\boldsymbol{\theta}}_{-i}[k]))]\\
&\leq V(\mathbf{a}_{-i}(\hat{\boldsymbol{\theta}}_{-i}[k]),\theta_i[k]),
\end{split}\end{align}
where we use the same logic as point (a) in \eqref{eq:positivetransfer}. 
\end{IEEEproof}
We remark that not participating in a frame is equivalent to free-riding, and our transfers ensure a lower cost is obtained when participating. However, as the net payment to the users is non-negative\footnote{While we don't prove it, we expect the transfer to be positive if the agent transmits, but we also note that it need not be zero if he doesn't, owing to the translation of viewpoints mentioned earlier.}, we will not immediately have budget-balance.  For the broader class of Bayes-Nash incentive-compatible mechanism, \cite{athey2013efficient} shows that only under the assumption of ``independent types" (the distribution of each agent's information is not directly affected by the other agents' information), budget can be balanced ex-interim.  However, in our system, each agent's information will have an impact on the other agents' information through the allocation.  Nevertheless, using the same technique of an initial sum being placed in escrow with the expectation that it would be returned at each stage (\textit{i.e,.} interim), our system may be budget-balanced.  Details using current prices of B2D service are provided in Section~\ref{sec:viability}.

\subsection{Value functions and optimal strategies}

We will now show that the value function given by the solution to \eqref{eq:valueMF} is well-defined and can be obtained using value iteration. Similarly, we will show that both the value function and the optimal allocation policy from a agent's perspective, given by \eqref{eq:valueAgentMF} and \eqref{eq:allocAgentMF} respectively, exist and can also be determined via value iteration.

Define operators $T_1$ and $T_2$ by
\begin{align}
& T_1{w}({\boldsymbol{\theta}})  =\sum_{i=1}^M v_i(\mathbf{a}^*({\boldsymbol{\theta}}),{\theta}_i) + \delta \E\left\{ {w}({\boldsymbol{\Theta}})\right\} \label{eq:recurMF}\\
\begin{split}
& T_2 \tilde{w}(1,(\theta_1,{\boldsymbol{\theta}}_{-1})) = \\
& \min_{\mathbf{a}} \sum_{i^\prime=1}^M v_{i^\prime}(\mathbf{a},\theta_{i^\prime}) + \delta \E\left\{\tilde{w}(1,(\Theta_1, {\boldsymbol{\Theta}}_{-1})) | \mathbf{a}, \theta_1\right\}
\end{split}
\label{eq:recurAgentMF}
\end{align}
using \eqref{eq:valueMF} and \eqref{eq:valueAgentMF}, respectively.
\begin{theorem}\label{thm:MFpolicy}
The following hold:
\begin{enumerate}
\item There exists a unique $W({\boldsymbol{\theta}})$ such that $T_1W({\boldsymbol{\theta}}) = W({\boldsymbol{\theta}})$, and given ${\boldsymbol{\theta}}$ for every $\mathbf{w}\in \mathbb{R}_+^M$, we have $\lim_{n\rightarrow\infty} T_1^n w=W({\boldsymbol{\theta}})$;
\item There exists a unique $\tilde{W}(1,(\theta_1,{\boldsymbol{\theta}}_{-1}))$ such that $T_2\tilde{W}(1,(\theta_1,{\boldsymbol{\theta}}_{-1})) = \tilde{W}(1,(\theta_1,{\boldsymbol{\theta}}_{-1}))$, and given $(\theta_1,{\boldsymbol{\theta}}_{-1})$ for every $\mathbf{w}\in \mathbb{R}_+^M$, we have $\lim_{n\rightarrow\infty} T_2^n w=\tilde{W}(1,(\theta_1,{\boldsymbol{\theta}}_{-1}))$; and
\item The Markov policy $\tilde{\mathbf{a}}((\theta_1,{\boldsymbol{\theta}}_{-1}))$ obtained from \eqref{eq:allocAgentMF} is an optimal policy to be used in cluster $j_1[\cdot]$ from the viewpoint of agent $1$.
\end{enumerate}
\end{theorem}
\begin{IEEEproof}
First, we consider statement 1). The proof follows by applying Theorem 6.10.4 in Puterman \cite{Put94}, and verifying the Assumptions 6.10.1, 6.10.2 and Propositions 6.10.1, 6.10.3.

Define the set of functions
\begin{equation}
\Phi=\left\{w:(\K,\T)^M \rightarrow \R^+: \sup_{\boldsymbol{\theta}\in (\K, \T)^M} \left|\frac{w(\boldsymbol{\theta})}{\alpha(\boldsymbol{\theta})}\right|<\infty\right\}
\end{equation}
where $\alpha(\boldsymbol{\theta})=\max\{\sum_i^M v_i(a^*(\boldsymbol{\theta}), \theta_i), 1\}$. Note that $\Phi$ is a Banach space with $\alpha$-norm,
\begin{equation}
||w||_{\alpha}=\sup_{\boldsymbol{\theta}\in (\K, \T)^M} \left|\frac{w(\boldsymbol{\theta})}{\alpha(\boldsymbol{\theta})}\right|<\infty
\end{equation}

Also define the operation $T_1$ as
\begin{equation}\small
T_1w(\boldsymbol{\theta})  =\sum_{i=1}^M v_i(\mathbf{a}^*(\boldsymbol{\theta}),{\theta}_i) + \delta \E\left\{ w(\boldsymbol{\Theta})\right\} 
\label{eq:recurMF}
\end{equation}
where $w\in \Phi$.

First, we need to show that for $\forall w\in \Phi$, $T_1w\in \Phi$. From Equation~(\ref{eq:recurMF}) and the definition of value functions, we know the sum of all users' values are bounded, say $\sum_{i=1}^M v_i(\mathbf{a}^*(\boldsymbol{\theta}),{\theta}_i)\leq A$. Then we have 
\begin{equation}
||T_1 w||_{\alpha} \leq A+\delta \E\left\{ w(\boldsymbol{\Theta})\right\} 
\end{equation}
where the rightside expression is bounded by the sum of $A$ and some multiple of $||w||_{\alpha}$. Hence, $T_1 w \in \Phi$.

Next, we need to verify Assumptions 6.10.1 and 6.10.2 in Puterman \cite{Put94}. Our theorem requires the verification of the following three conditions. Let $\boldsymbol{\Theta}[k]$ be the random variable denoting the current system state at frame $k$, where $\boldsymbol{\Theta}[k]=(\boldsymbol{d}[k-1], \boldsymbol{e}[k])$. Then we must show that $\forall \theta \in (\K,\T)^M$, for some constants $0<\gamma_1<\infty$, $0<\gamma_2<\infty$ and $0<\gamma_3<1$,
\begin{equation}
\sup_{a\in A} |\sum_i^M v_i(a^*(\boldsymbol{\theta}), \theta_i)|\leq \gamma_1 \alpha(\theta)
\label{eq:cond1}
\end{equation}
\begin{equation}
\mathbb{E}_{\boldsymbol{\theta}[1]} [w(\boldsymbol{\theta}[1]) | \boldsymbol{\theta}[0]=\theta]\leq \gamma_2 \alpha(\theta), \quad \forall w\in \Phi
\label{eq:cond2}
\end{equation}
\begin{equation}
{\beta}^k \mathbb{E}_{\boldsymbol{\theta}[k]} [\alpha(\boldsymbol{\theta}[k]) | \boldsymbol{\theta}[0]=\theta]\leq \gamma_3 \alpha(\theta), \quad \text{for some k}
\label{eq:cond3}
\end{equation}

~(\ref{eq:cond1}) holds from the definition of $\alpha(\boldsymbol{\theta})=\max\{\sum_i^M v_i(a^*(\boldsymbol{\theta}), \theta_i), 1\}$.

~(\ref{eq:cond2}) holds true since
\begin{equation}
\begin{aligned}
\mathbb{E}_{\boldsymbol{\theta}[1]} [w(\boldsymbol{\theta}[1]) | \boldsymbol{\theta}[0]&=\theta]\leq ||w||_{\alpha}\times \mathbb{E}_{\boldsymbol{\theta}[1]} [\alpha(\boldsymbol{\theta}[1]) | \boldsymbol{\theta}[0]=\theta]\\
&\leq ||w||_{\alpha} \times {\gamma}'_2\alpha(\theta), \quad \text{for some large enough ${\gamma}'_2$}\\
&=\gamma_2 \times \alpha(\theta)
\end{aligned}
\label{eq:cond2:proof}
\end{equation}
as we know in our mean field model, $\boldsymbol{\theta}[1]$ are all drawn i.i.d. from the given distribution $[\otimes \rho^M, \otimes \zeta^M],$ with $\rho$ pertaining to the deficit, and $\zeta$ pertaining to the B2D transmissions received by that agent, so the first inequality holds in~(\ref{eq:cond2:proof}).

Finally, we have~(\ref{eq:cond3}) since, 
\begin{equation}
\begin{aligned}
&{\beta}^k \mathbb{E}_{\boldsymbol{\theta}[k]} [\alpha(\boldsymbol{\theta}[k]) | \boldsymbol{\theta}[0]=\theta]\\
&={\beta}^k \mathbb{E}_{\boldsymbol{\theta}[k]} [\sum_i^M v_i(a^*(\boldsymbol{\theta}[k]), \theta_i)|\boldsymbol{\theta}[0]=\theta]\\
&\leq {\beta}^j\times {\gamma}'_3 \alpha(\boldsymbol{\theta})\\
&=\gamma_3 \alpha(\boldsymbol{\theta})
\end{aligned}
\label{eq:cond3:proof}
\end{equation}
The first equality holds from the definition of $\alpha(\theta)$, and the first inequality holds true is because in our mean field mode, $\boldsymbol{\theta}[j]$ are all drawn i.i.d. from the given distribution $[\otimes \rho^M, \otimes \zeta^M],$ with $\rho$ pertaining to the deficit, and $\zeta$ pertaining to the B2D transmissions received by that agent, so it's identical for all $k$.

Since we have verified all the three conditions required by Theorem 6.10.4 in Puterman, Statement 1) holds true.

For statement 2), we can use the same argument as the above proof to show the existence of fix point. We omit the details here. The last part of Theorem~\ref{thm:MFpolicy} follows from the discussion before the statement of this theorem.
\end{IEEEproof}

\section{Mean Field Equilibrium}\label{sec:mfe-1}

In the mean-field setting, assuming the state of every other agent is drawn \emph{i.i.d.} with distribution $\rho \times \zeta$, the deficit of any given agent evolves as a Markov chain. We start by showing that this Markov chain has a stationary distribution. If this stationary distribution is the same as $\rho$, then the distribution $\rho$ is defined as a mean-field equilibrium (MFE); we use the Schauder fixed point theorem to show the existence of a fixed point $\rho$. Using the regenerative representation of the stationary distribution of deficits given $\rho$ and a strong coupling result, we prove that the mapping that takes $\rho$ to the stationary distribution of deficits is continuous using a strong coupling result. Finally, we show that the set of probability measures to be considered is convex and compact so that existence follows. 

\subsection{Stationary distribution of deficits}
Fix a typical agent $i$ and consider the state process $\{d_i[k]\}_{k=-1}^\infty$. This is a Markov process in the mean-field setting: if there is no regeneration, then the deficit changes as per the allocation and the number of B2D packets received, and is chosen via the regeneration distribution otherwise. The allocation is a function of the past $d_i$, the number B2D packets received and the state of the other agents. The number of B2D packets received and the state of the other agents are chosen $i.i.d$ in every frame. This Markov process has an invariant transition kernel. We construct it by first presenting the form given the past state and the allocations, namely,
\begin{equation}
\begin{aligned}
&\pr(d_i[k]\in B| d_i[k-1]=d, e_i[k]=e, \mathbf{a})\\
&= \delta 1_{\big\{\big(d+\eta_i-\chi_i(\mathbf{a},(d,e))\big)^+\in B\big\}} + (1-\delta) \Psi(B),
\end{aligned}
\label{equ:transitionprob1}
\end{equation}
where $B\subseteq \R^+$ is a Borel set and $\Psi$ is the density function of the regeneration process for deficit. In the above expression, the first term corresponds to the event that agent $i$ can either decode the packet using D2D transmissions or not, and the second term captures the event that the agent regenerates after frame $k$. Using \eqref{equ:transitionprob1} we can define the one-step transition kernel $\tilde{\Upsilon}$ for the Markov process as
\begin{align}
\begin{split}
& \tilde{\Upsilon}(B,d)=\pr(d_i[k]\in B| d_i[k-1]=d) \\
& = \delta \int 1_{\big\{\big(d+\eta_i-\chi_i(\mathbf{a}^*((d,e),\hat{\boldsymbol{\theta}}_{-i}),(d,e))\big)^+\in B\big\}}\\
&\times d(\otimes\rho^{M-1}\times\otimes\zeta^{M-1})(\hat{\boldsymbol{\theta}}_{-i})d\zeta(e)  + (1-\delta) \Psi(B).
\end{split}
\label{equ:transitionprob}
\end{align}
For later use we also define the transition kernel without regeneration but one obtained by averaging the states of the other users while retaining the state of user $i$, i.e.,
\begin{align}
\begin{split}
& \Upsilon(B|d,e)=\\
& \pr(d_i[k]\in B| \text{ no regeneration}, d_i[k-1]=d, e_i[k]=e) \\
& =  \int 1_{\big\{\big(d+\eta_i-\chi_i(\mathbf{a}^*((d,e),\hat{\boldsymbol{\theta}}_{-i}),(d,e))\big)^+\in B\big\}}\\
& \qquad \times d(\otimes\rho^{M-1}\times\otimes\zeta^{M-1})(\hat{\boldsymbol{\theta}}_{-i})d\zeta(e)  
\end{split}
\label{equ:txprob2}
\end{align}
The $k$ fold iteration of this transition kernel is denoted by ${\Upsilon}^{(k)}$.

\begin{lemma}\label{lem:stationary}
The Markov chain where the allocation is determined using \eqref{eq:allocMF} based on choosing the states of all users other than $i$ \emph{i.i.d.} with distribution $\rho\times\zeta$ and the number of B2D packets of user $i$ independently with distribution $\zeta$, and the transition probabilities in~(\ref{equ:transitionprob1}) is positive Harris recurrent and has a unique stationary distribution. We denote the unique stationary distribution for the deficit of a typical agent by $\Pi_{\rho\times \zeta}$; the dependence on $\Psi$ is suppressed. The expression of this stationary distribution $\Pi_{\rho\times \zeta}$ in term of $\Upsilon^{(k)}_{\rho\times\zeta}(B|D,E)$ is given as, 
\begin{align}
\Pi_{\rho\times\zeta}(B)=\sum_{k=0}^\infty (1-\delta) \delta^k \E_{\Psi}(\Upsilon^{(k)}_{\rho\times\zeta}(B|D,E))
\end{align}
where $D=\{D_k\}_{k\in\mathbb{N}}$ is the deficit process, $E=\{E_k\}_{k\in\mathbb{N}}$ is the B2D packet reception process, and $\E_{\Psi}(\Upsilon^{(k)}_{\rho\times\zeta}(B|D,E)) = \int \Upsilon^{(k)}_{\rho\times\zeta}(B | d,e) d\Psi(d) d\zeta(e)$.
\end{lemma}
\begin{IEEEproof}
First, from~(\ref{equ:transitionprob1}), we note the Doeblin condition, namely,
\begin{equation}
\hspace{-0.01in}\pr(d_i[k]\in B| d_i[k-1]=d, e_i[k]=e, \mathbf{a})\geq (1-\delta)\Psi(B)
\end{equation}
where $0<\delta<1$ and $\Psi$ is a probability measure. Then following the results in Chapter 12 of \cite{MeyTwe09}, the Markov chain with transition probabilities in~(\ref{equ:transitionprob1}) is positive Harris recurrent and has a unique stationary distribution.

Next, let $-\tau$ be the last time before $0$ that regeneration happened.  We have 
\begin{align}
\begin{split}
\Pi_{\rho\times\zeta}(B)&=\sum_{k=0}^{\infty}\pr(B, \tau=k)\\
                                     &=\sum_{k=0}^{\infty}\pr(B|\tau=k)\pr(\tau=k)
\end{split}
\end{align}
Since the regeneration happens independently of the deficit queue with inter-regeneration times geometrically distributed with parameter $(1-\delta)$, it follows that $\pr(\tau=k)=(1-\delta){\delta}^k$. Hence
\begin{align}
\begin{split}
&\Pi_{\rho\times\zeta}(B)=\sum_{k=0}^{\infty}(1-\delta){\delta}^k\pr(D[0]\in B|\tau=k)\\
                                     &=\sum_{k=0}^{\infty}(1-\delta){\delta}^k\E(\E(1_{\{D[0]\in B\}}|\tau=k, D_{-k}=D, E)|\tau=k)\\
                                     &=\sum_{k=0}^{\infty}(1-\delta){\delta}^k\E(\Upsilon^{(k)}_{\rho\times\zeta}(B|D,E)|\tau=k)\\
                                     &=\sum_{k=0}^{\infty}(1-\delta){\delta}^k\E_{\Psi}(\Upsilon^{(k)}_{\rho\times\zeta}(B|D,E))
\end{split}
\end{align}
where the last equality holds since $D_{-k}\sim\Psi$ given $\tau=k$.
\end{IEEEproof}

\subsection{Agent and cluster decision problems}
Suppose that each agent has common information about the distribution for the deficit $\rho\in \mathcal{M}_1(\mathbb{K})$ (where $\mathcal{M}_1(\K)$ is the set of probability measures on $\K$); this is one of the mean-field assumptions. We further assume that $\rho \in \Pp$ where
\begin{equation}\small
\hspace{-0.025in} \Pp=\big\{\rho|\rho\in\mathcal{M}_1(\K) \text{ with finite mean}\big\}.
\end{equation}
We will also assume that the regeneration distribution $\Psi\in \Pp$. From Section~\ref{section2}, the best strategy for each agent is to truthfully reveal its state based on the transfers suggested in each frame as per \eqref{eq:transferMF}. Then each cluster simply maximizes the system value function by choosing the greedy optimal allocation based on \eqref{eq:allocMF}.

\subsection{Mean field equilibrium}
Given the distribution for deficit $\rho$ and the station distribution $\Pi_{\rho\times \zeta}$, we have the following definition.
\begin{definition}
(Mean field equilibrium). Let $\rho$ be the common cumulative distribution for deficit and telling-truth is the optimal policy for each agent in every frame. Then, we say that the given $\rho$ along with the truth-telling behavior constitutes a mean field equilibrium if 
\begin{equation}
\rho(d)=\Pi_{\rho\times \zeta}(d),  \forall d\in \K
\end{equation}
\end{definition}

\section{Existence of MFE}\label{sec:mfe-2}
The main result showing the existence of MFE is as follows.
\begin{theorem}\label{thm:MFexistence}
There exists an MFE of $\rho$ and truth-telling policy such that 
$\rho(d)=\Pi_{\rho\times \zeta}(d)$,  $\forall d\in \K$.
\end{theorem}

As mentioned earlier, we will be specializing to the space $\mathcal{M}_1(\K)$, its subset $\Pp$ and further subsets of $\Pp$. The primary topology on $\mathcal{M}_1(\K)$ that we will consider is the uniform norm topology, i.e., using the $l_\infty$ norm  given by $\|\rho \|=\max_{d \in \K} \rho(d)$. Another topology on $\mathcal{M}_1(\K)$ that we will use is the point-wise convergence topology, i.e., $\{\rho_n\}_{n=1}^\infty \subset \mathcal{M}_1(\K)$ converges to $\rho \in \mathcal{M}_1(\K)$ point-wise if $\lim_{n\rightarrow \infty} \rho_n(d)=\rho(d)$ for all $d\in \K$; it is easily verified that the convergence is the same as weak convergence of measures. Also, define the mapping $\Pi^*$ that takes $\rho$ to the invariant stationary distribution $\Pi_{\rho\times \zeta}(\cdot)$. Let $\Pp^\prime \subset \Pp$. We will use the Schauder fixed point theorem to prove existence which is given as follows. 
\begin{theorem}\label{thm:SFP}
(Schauder Fixed Point Theorem). Suppose $\F(\Pp^\prime)\subset \Pp^\prime$, $\F$ is continuous and $\F(\Pp^\prime)$ is contained in a convex and compact subset of $\Pp^\prime$,  then $\F$ has a fixed point.
\end{theorem}

Note that from the definition of $\Pp$, it is already convex. Then in the following section, we will prove that under the topology generated by the uniform norm, $\Pi^*$ is continuous and the image of $\Pi^*$ for a specific subset $\Pp^\prime$ is pre-compact.

\subsection*{Steps to Prove  MFE Existence}\label{sec:mfe-proof}
We first need to prove the continuity of $\Pi^*$ with the uniform norm topology. For this we will start by showing that for any sequence $\rho_n\rightarrow \rho$ with $\rho_n, \rho \in \Pp$ in uniform norm, $\Pi^*(\rho_n)\Rightarrow \Pi^*(\rho)$ 
(where $\Rightarrow$ denotes weak convergence). Finally, using some properties of $\mathcal{M}_1(\K)$ we will strengthen the convergence result to prove that $\Pi^*(\rho_n)\rightarrow \Pi^*(\rho)$ in uniform norm too.

\subsubsection{Continuity of the mapping $\Pi^*$}
We will restrict our attention to subset of probability measures $\Pp(F) \subset \mathcal{M}_1(\K)$ such that
\begin{align}
\Pp(F) = \left\{ \rho \in \mathcal{M}_1(\K): \sum_{d\in \K} d \rho(d) \leq F \right\}
\end{align}
where $F$ is a given non-negative constant; in other words, probability measures with a specified bound on the mean and not just a finite mean. We will assume that the regeneration distribution $\Psi\in \Pp(F^\prime)$ for some $F^\prime$. Later on we will specify the values of $F$ and $F^\prime$ to be used. 

We start with the following preliminary result that establishes compactness of sets like $\Pp(F)$ in the uniform norm topology; note that convexity is immediate. 
\begin{lemma}\label{lem:compactlinf}
Given a sequence of non-negative numbers $\{b_n\}_{n\in \mathbb{N}}$ such that $\lim_{n\rightarrow \infty} b_n = 0$, then ${\cal C}=\big\{ x: |x_n| \leq b_n \; \forall n\in\mathbb{N} \big\}$ is a compact subset of $l_\infty$ and sequences of elements from ${\cal C}$ that converge point-wise also converge uniformly.
\end{lemma}
\begin{IEEEproof}
We will establish the second property first. We're given a sequence $\{\sigma^n\}_{n\in\mathbb{N}} \subset {\cal C}$ that converges point-wise to $\sigma$; it obviously follows that $\sigma\in {\cal C}$ even with point-wise convergence so that we are, in fact, showing that ${\cal C}$ is closed in $l_\infty$ too. Since $\lim_{n\rightarrow\infty} b_n = 0$, given $\epsilon > 0$, there exists~\footnote{Note the abuse of notation only in this section to use $N$ to represent a positive integer.} $N$ such that for all $n > N$, $b_n \leq \epsilon/2$ so that $\sup_{k\in\mathbb{N}} |\sigma_n^k| \leq b_n \leq \epsilon/2$ too. Since $\lim_{k\rightarrow\infty} \sigma^k_n = \sigma_n$ for all $n=1,\dotsc,N$, we can find $N_n$ such that for all $k > N_n$, $|\sigma^k_n - \sigma_n| \leq \epsilon$. Therefore, for $k > \max(N,\max_{n=1,\dotsc,N} N_n)$
\begin{align}
| \sigma_n^k -\sigma_n | \leq 
\begin{cases}
\epsilon & n = 1, \dotsc, N\\
| \sigma_n^k| + |\sigma_n| \leq \epsilon & n > N
\end{cases}
\end{align}
so that $\| \sigma^k - \sigma \| \leq \epsilon$.

Since we have already established that ${\cal C}$ is closed in $l_\infty$, it is sufficient to prove that it is totally bounded as well. Here we first find $N$ such that for all $n > N$, $b_n \leq \epsilon$ so that $\sup_{k\in\mathbb{N}} |\sigma_n^k| \leq b_n \leq \epsilon$ too. Then from the compactness of $\prod_{n=1}^N [-b_n, b_n]\in \mathbb{R}^N$, we can find a finite number of points $\{ v^1, v^2, \dotsc, v^L \} \subset \prod_{n=1}^N [-b_n, b_n]$ such that $\prod_{n=1}^N [-b_n, b_n]$ is covered by balls of radius $\epsilon$ around $v^l$, $l=1,\dotsc,L$. Now we construct $\{\hat{v}^1, \dotsc, \hat{v}^L\}\in {\cal C}$ as follows for $l=1,\dotsc,L$
\begin{align}
\hat{v}^l_n = 
\begin{cases}
v^l_n & \text{ if } n \leq N \\
0 & \text{otherwise}
\end{cases}
\end{align}
By our choice of $N$, $\{\hat{v}^1, \dotsc, \hat{v}^L\}$ is a finite cover of ${\cal C}$ with balls of radius $\epsilon$, proving that ${\cal C}$ is totally bounded too.
\end{IEEEproof}
One can also use the Cantor diagonalization procedure to show sequential compactness in the proof above.

We have an immediate corollary of this result.
\begin{corollary}\label{lem:compactPe}
The set of probability measures $\Pp(F)$ on $\K$ is a compact set of $l_\infty$ for every $F\in \R_+$.
\end{corollary}
\begin{IEEEproof}
For any $\rho \in \Pp(F)$, $p(d_1) \leq 1$ and by Markov's inequality for $n > 1$
\begin{align}
p(d_n) \leq \sum_{k=n}^\infty p(d_k) \leq \frac{F}{d_n}
\end{align}
with $\lim_{n\rightarrow\infty} \tfrac{F}{d_n}=0$. Using Lemma~\ref{lem:compactlinf} the result follows.
\end{IEEEproof}

Next, we present a coupling result from Thorisson \cite[Theorem 6.1, Chapter 1]{Thorisson00}. This result will be used in proving continuity of the stationary distribution of the deficit process under the topology of point-wise convergence and in strengthening the convergence result.
\begin{theorem}\label{thm:coupling}
Let $\{ \rho_n \}_{n=1}^{\infty}\in \mathcal{M}_1(\K)$ converge weakly to $\rho \in \mathcal{M}_1(\K)$, then there exists a coupling, i.e., random variables $\{X_n\}_{n=1}^{\infty}$, $X$ on a common probability space and a random integer $N$ such that $X_n\sim \rho_n$ for all $n\in \mathbb{N}$, $X \sim \rho$ and $X_n = X$ for $n \geq N$.
\end{theorem}

This result shows that weak convergence of probability measures on $\K$ is equivalent to convergence of probability measures in total variation norm, and hence, also in uniform norm.

Next we show that $\Pi_{\rho \times \zeta} \in \Pp(F)$ whenever $\rho \in \Pp(F)$. 
\begin{lemma}
If $\rho \in \Pp(F)$ for $F\geq \tfrac{\delta \eta}{1-\delta}$ and the regeneration distribution $\Psi\in \Pp(F^\prime)$ for $F^\prime \leq F-\tfrac{\delta \eta}{1-\delta}$, then the stationary distribution of the deficit process of any specific user $\Pi_{\rho\times \zeta} \in \Pp(F)$.
\end{lemma}
\begin{IEEEproof}
The proof will involve three steps. The first is to establish that $\Pi_{\rho\times \zeta}$ is indeed a probability distribution, which is obvious. The second is to establish that $\Pi_{\rho\times \zeta} \in \mathcal{M}_1(\K)$, which will be carried out using induction by analyzing the properties of the Markov transition kernel of the deficit process without any regenerations. Finally, using stochastic dominance we will show that $\Pi_{p\times \zeta} \in \Pp(F)$.

From earlier Lemma~\ref{lem:stationary}, we know that 
\begin{align}
\Pi_{\rho\times\zeta}(B)=\sum_{k=0}^\infty (1-\delta) \delta^k \E_{\Psi}(\Upsilon^{(k)}_{\rho\times\zeta}(B|D,E))
\end{align}
Therefore, for our proof we will show that $\E_{\Psi}(\Upsilon^{(k)}_{\rho\times\zeta}(\cdot|D,E))\in \mathcal{M}_1(\K)$. Since 
\begin{align}
\E_{\Psi}(\Upsilon^{(k)}_{\rho\times\zeta}(B|D,E)) = \int \Upsilon^{(k)}_{\rho\times\zeta}(B | d,e) d\Psi(d) d\zeta(e),
\end{align}
and $\Psi\in \mathcal{M}_1(\K)$ and $\zeta \in \mathcal{M}_1(\T)$, it is sufficient to show that $\Upsilon^{(k)}_{\rho\times\zeta}(\cdot | d,e) \in \mathcal{M}_1(\K)$ for every $(d,e) \in (\K, \T)$.

Since $\Upsilon^{(0)}_{\rho\times\zeta}(\cdot | d,e)$ is a point-mass at $d$, the initial condition is satisfied. We now make the induction assumption that $\Upsilon^{(k)}_{\rho\times\zeta}(\cdot | d,e) \in \mathcal{M}_1(\K)$ and show that this implies that $\Upsilon^{(k+1)}_{\rho\times\zeta}(\cdot | d,e) \in \mathcal{M}_1(\K)$. Since $\Upsilon^{(k+1)}_{\rho\times\zeta}(\cdot | d,e)$ is a probability measure, we only need to show that its support is $\K$. By the definition of the Markov transition kernel without regenerations, we have
\begin{equation}
\begin{aligned}
&\Upsilon^{(k+1)}_{\rho\times\zeta}(B | d,e) \\
&= \int \sum_{j=0}^1 1_{ \{(d^\prime+\eta-j)_+ \in B\} } p_j(d^\prime, e^\prime) d\Upsilon^{(k)}(d^\prime | d, e) d\zeta(e^\prime)
\end{aligned}
\end{equation}
for some measurable functions $\{p_j(d^\prime, e^\prime)\}_{j=0,1}$ that account for the states of the other users being chosen independently using distribution $\rho\times\zeta$ and the greedy optimal allocation function $\mathbf{a}^*(\cdot)$. The assertion that $\Upsilon^{(k+1)}_{\rho\times\zeta}(\cdot | d,e) \in \mathcal{M}_1(\K)$ follows since $d^\prime \in \K$ and the only possible updates are an increase of the deficit to $d^\prime+\eta$ or a decrease to either $0$ or $d^\prime+\eta -1$ (depending on value of $d^\prime$).

The deficit process for any given user is stochastically dominated by the fictitious process where the user is never allowed to decode the contents of a frame during his lifetime, this is irrespective of his state or the state of the other users. Denote this process by $\{ \tilde{D}_k \}_{k\in \mathbb{N}}$; it is easily discerned that the process takes values in $\K$. The transition kernel for this process is given by
\begin{align}
\mathbb{P}(\tilde{D}_{k+1}=d | \tilde{D}_{k}=d^\prime) = \delta 1_{\{d=d^\prime+\eta\}} + (1-\delta) \Psi(d).
\end{align}
Using the same proof as in Lemma~\ref{lem:stationary}, the invariant distribution $\tilde{\Pi}$ of the $\{ \tilde{D}_k \}_{k\in \mathbb{N}}$ process is given by
\begin{align}
\tilde{\Pi}(d) = \sum_{k=0}^\infty (1-\delta) \delta^k \sum_{d^\prime \in \K} \Psi(d^\prime) 1_{\{d^\prime+k\eta=d\}}.
\end{align}
By the stochastic ordering property, the proof follows by noting that
\begin{equation}
\begin{aligned}
\E_{\Pi_{\rho\times\zeta}}[D] &\leq \E_{\tilde{\Pi}}[D] \\
& \leq \sum_{k=0}^\infty (1-\delta) \delta^k \sum_{d^\prime \in \K} \Psi(d^\prime) (d^\prime+k \eta)\\
& < F^\prime + \frac{\delta\eta}{1-\delta}
\end{aligned}
\end{equation}
\end{IEEEproof}

Next we show continuity properties of the mapping $\Pi^*$.
\begin{theorem}
The mapping $\Pi^*: \Pp(F) \mapsto \Pp(F)$ is continuous in the uniform topology. In addition, $\Pi^*$ has a fixed point in $\Pp(F)$.
\end{theorem}
\begin{IEEEproof}
We will start by showing that $\Pi^*$ is continuous in the topology of point-wise convergence. For this we will use the coupling from Theorem~\ref{thm:coupling} to establish convergence in total variation norm of the Markov transition kernels of the deficit process without any regenerations. Then using Lemma~\ref{lem:compactlinf} we can strengthen the topology to complete the proof of the first part. The fixed point result then follows from the Schauder fixed point theorem after noting both the convexity and compactness of $\Pp(F)$.

To establish the continuity of $\Pi^*$ in the topology of point-wise convergence, we will start by proving that the Markov transition kernels without regeneration $\{\Upsilon_{\rho\times\zeta}^{(k)}(\cdot | d, e)\}_{k=0}^\infty$ are continuous in the topology of point-wise convergence. Since $\Upsilon_{\rho\times\zeta}^{(0)}(\cdot | d, e)$ is a point-mass at $d$ irrespective of $\rho\in \Pp(F)$, the continuity assertion holds. In fact, for all $n \geq 1$ and $d^\prime \in \K$, $\Upsilon_{\rho_n\times\zeta}^{(0)}(d^\prime | d, e)=\Upsilon_{\rho\times\zeta}^{(0)}(d^\prime | d, e)$. Let $\{ \rho_n \}_{n\in\mathbb{N}} \subset \Pp(F)$ be a sequence converging point-wise~\footnote{By Lemma~\ref{lem:compactlinf}, this convergence also holds in $l_\infty$.} to $\rho\in \Pp(F)$. We will show that $\Upsilon_{\rho_n\times\zeta}^{(k)}(\cdot | d, e)$ converges point-wise to $\Upsilon_{\rho\times\zeta}^{(k)}(\cdot | d, e)$ for all $k\in \mathbb{N}$. We will prove this by induction.

We will refer to any measures and random variables corresponding to $\rho_n$ as coming from the $n^{\mathrm{th}}$ system and those corresponding to $\rho$ as coming from the limiting system. We will prove the point-wise convergence of $\Upsilon_{\rho_n\times\zeta}^{(k)}(\cdot | d, e)$ converges point-wise to $\Upsilon_{\rho\times\zeta}^{(k)}(\cdot | d, e)$ for all $k\in \mathbb{N}$ using the metric given by the total variation norm. Following Lindvall \cite{Lindvall92}, the total variation norm distance between two probability measures $\mu$ and $\nu$ on a countable measurable probability space $\Omega$ is given by
\begin{equation}
\begin{aligned}
&d_{TV}(\mu,\nu)  = \frac{1}{2} \sum_{\omega \in \Omega} \big| \mu(\omega) - \nu(\omega) \big|\\
& =\inf\{ \pr(X\neq Y): \text{r.v.s } X, Y \text{s.t. } X\sim \mu \text{ and } Y\sim \nu\},
\end{aligned}
\end{equation}
where the infimum is over all couplings or joint distributions such that the marginals are given by $\mu$ and $\nu$, respectively; the second definition applies more generally while the first is restricted to countable spaces. 

For ease of exposition we will denote by $1$ the user whose deficit varies as per the Markov transition kernel $\Upsilon_{\bullet\times\zeta}^{(k)}(\cdot | d, e)$ and the remaining users in the cluster by indices $\{2, 3, \dotsc, M\}$. For the $n^{\mathrm{th}}$ system and in the limiting system, in every frame the B2D component of the state of every user (including $1$) is chosen $\emph{i.i.d.}$ with distribution $\zeta$. We will couple all the systems under consideration such that the B2D component of the state is exactly the same; denote the random vector by $\mathbf{E}$ with components $E_l$ for $l\in \{1, 2, \dotsc, M\}$. For users $l\in \{2, 3, \dotsc, M\}$ the deficit is chosen independently via distribution $\rho_n$ in the $n^{\mathrm{th}}$ system and via distribution $\rho$ in the limit system. Since $\rho_n$ converges to $\rho$ point-wise, using Theorem~\ref{thm:coupling} we can find a coupling $\{ \tilde{X}_n^l\}_{n\in \mathbb{N}}$, $\tilde{X}^l$ and an \emph{a.s.} finite random integer $\tilde{N}_l$ for $l\in \{2, 3, \dotsc, M\}$ such that for $n\geq \tilde{N}^l$, $\tilde{X}_n^l = \tilde{X}^l$. 

Next by the induction hypothesis let $\Upsilon_{\rho_n\times\zeta}^{(k)}(\cdot | d, e)$ converge point-wise to $\Upsilon_{\rho\times\zeta}^{(k)}(\cdot | d, e)$ for some $k \in \mathbb{N}$, once again by Theorem~\ref{thm:coupling}, there exists a coupling $\{ X_n \}_{n\in \mathbb{N}}$, $X$ and an \emph{a.s.} finite random variable $N_k \in \mathbb{N}$ such that $X_n \sim \Upsilon_{\rho_n\times\zeta}^{(k)}(\cdot | d, e)$ for all $n \in \mathbb{N}$, $X\sim \Upsilon_{\rho\times\zeta}^{(k)}(\cdot | d, e)$ and $X_n=X$ for all $n \geq N_k$. 

With these definitions in place, further define the following
\begin{equation}
\begin{aligned}
& D_n^{k+1}  =  \\
& \Bigg(X_n + \eta - \chi_1\bigg(\mathbf{a}^*\Big(\big( (X_n, E_1), (\tilde{X}_n^2, E_2), \dotsc, (\tilde{X}_n^M, E_M)\big)\Big),  \\
&\Big(\big( (X_n, E_1), (\tilde{X}_n^2, E_2), \dotsc, (\tilde{X}_n^M, E_M)\big)\Big)\bigg) \Bigg)_+ 
\end{aligned}
\end{equation}
\begin{equation}
\begin{aligned}
& D^{k+1}  =  \\
& \Bigg(X + \eta - \chi_1\bigg(\mathbf{a}^*\Big(\big( (X, E_1), (\tilde{X}^2, E_2), \dotsc, (\tilde{X}^M, E_M)\big)\Big),  \\
&\Big(\big( (X, E_1), (\tilde{X}^2, E_2), \dotsc, (\tilde{X}^M, E_M)\big)\Big)\bigg) \Bigg)_+,
\end{aligned}
\end{equation}
where we have taken care to explicitly spell out the states of all the users involved.

Then $D_n^{k+1}$ is a random variable distributed as $\Upsilon_{\rho_n\times\zeta}^{(k+1)}(\cdot | d, e)$ and $D^{k+1}$ is a random variable distributed as $\Upsilon_{\rho\times\zeta}^{(k+1)}(\cdot | d, e)$. Furthermore, for $n \geq \hat{N}:=\max(N_k, \tilde{N}_2,\dotsc, \tilde{N}_M)$, we have $X_n=X$, $\tilde{X}_n^l=\tilde{X}^l$ for $l\in \{2, 3, \dotsc, M\}$. The last statement then implies that $D_n^{k+1}=D^{k+1}$ for $n \geq \hat{N}$. Therefore, it follows that 
\begin{align}
\{ \omega: D_n^{k+1} \neq D^{k+1} \} \subset \{ w: \hat{N} > n \},
\end{align}
so that
\begin{equation}
\begin{aligned}
d_{TV}\Big(\Upsilon_{\rho_n\times\zeta}^{(k+1)}(\cdot | d, e), \Upsilon_{\rho\times\zeta}^{(k+1)}(\cdot | d, e)\Big) &\leq \mathbb{P}\big(D_n^{k+1}\neq D^{k+1}\big) \\
&\leq \mathbb{P}( \hat{N} > n )
\end{aligned}
\end{equation}
which converges~\footnote{Note that this yields a rate of convergence result as well.} to $0$ as $n\rightarrow\infty$ by the \emph{a.s.} finiteness of $\hat{N}$. From the definition of the metric $d_{TV}(\cdot,\cdot)$, it is follows that $\Upsilon_{\rho_n\times\zeta}^{(k+1)}(\cdot | d, e)$ converges to $\Upsilon_{\rho\times\zeta}^{(k+1)}(\cdot | d, e)$ in $l_1$, and so both in $l_\infty$ and point-wise also.

Having established the basic convergence result, $\E_{\Psi}(\Upsilon^{(k)}_{\rho_n\times\zeta}(\cdot |D,E))$ converges point-wise to $\E_{\Psi}(\Upsilon^{(k)}_{\rho\times\zeta}(\cdot |D,E))$ for every $k\in \{0\}\cup \mathbb{N}$ by using the bounded convergence theorem since we are averaging probability distributions. Additionally, again using the bounded convergence theorem, $\Pi_{\rho_n\times\zeta}(\cdot)$ converges point-wise to $\Pi_{\rho\times\zeta}(\cdot)$.
\end{IEEEproof}

\begin{theorem}\label{thm:MEFuniqueness}
The MFE is unique.
\end{theorem}
\begin{IEEEproof}
Suppose there exist two MFE, namely $\rho_1$ and $\rho_2$. 
Consider a generic agent $1$.  Agent $1$ has a belief that the other agents in the same cluster will draw their states from $\rho_1$ or $\rho_2$ for deficits and $\zeta$ for B2D transmissions in an i.i.d. fashion. We assume that each agent has the same realization of B2D packets received under these two deficit distributions. Given this belief and our incentive compatible mechanism (that determines transfers as a fucntion of the belief),  all the agents in this cluster will truthfully reveal their states, \emph{i.e.}, the B2D term will be the same no matter whether the belief is $\rho_1$ or $\rho_2$.  By Algorithm~\ref{alg: D2D}, this will result in the same deficit update for agent $1$.  Therefore, given the truth-telling mechanism and the unique policy, we achieve a unique MFE, i.e., $\rho_1=\rho_2.$
\end{IEEEproof}

We end with a few remarks on generalizing the D2D transmission model. As mentioned earlier, we constrain our analysis to the case of D2D transmissions being error-free even though errors can occur in practice. Also, in our implementation we follow the WiFi distributed coordination function (DCF). We will now describe at a high-level how both the incentive mechanism and the existence of the mean-field equilibrium hold under both these scenarios. We will start with the incentive mechanism. Here the exact same logic holds, except that the transition kernels for the user deficits are much more involved so that calculating the system-optimum policy and value function from the user's perspective becomes harder. Next we discuss whether a mean-field equilibrium exists or not. The main technical challenge again would be to show the continuity properties to apply the Schauder fixed point theorem. For the scenario with D2D transmissions with errors, we will endow each of the user with $T$ error sequences\footnote{Each of these error sequences would list which of the other users receive the D2D chunk.}  for each frame corresponding to the maximum possible transmissions to be made by the user. For the purposes of demonstrating continuity, we will then couple systems by insisting on these error sequences being exactly the same. For the scenario with the WiFi DCF, we will endow each user with $T$ countdown timer values, again corresponding to the maximum possible transmissions made by the user. Again we will couple systems by insisting on these countdown timers being exactly the same. With the two coupling ideas in the place, we can then use the same ideas as our proof above to show the continuity and the existence of a mean-field equilibrium. Note that we can also combine D2D transmissions with the WiFi DCF. We omit the exact details of these proofs in the interests of brevity.

\section{Passage to the Mean Field Limit}\label{sec:passage}
We gave an overview of the finite agent system in Sections~\ref{sec:PBE-MFE} (description of FiniteDPM) and \ref{sec:system-model}.  Here, we briefly discuss the passage between the finite agent system and the mean field model that we have used throughout the paper. As in other literature on repeated games under the mean field setup \cite{IyeJoh14,ManRam14}, we have considered the system with an infinitely large number of agents at finite time.  It is straight-forward to follow the steps in \cite{IyeJoh14,ManRam14} to prove convergence of the finite agent system to the mean-field model in our context.  However, to the best of our knowledge, the study of mean field games as time also becomes infinitely large is currently open.  There has been recent work in non-game-theoretical settings (using a fixed policy) studying the question of the conditions required to ensure that the mean field model is indeed the limiting case of the finite system when time becomes asymptotically large \cite{BenBou08,BorSun12}. In the case of our system, the set of measures that we consider is tight, since they are all stochastically dominated by a fictitious system in which no D2D transmissions happen and the agents' deficits simply increase and then they regenerate.  Furthermore, we showed in Theorem~\ref{thm:MEFuniqueness} that the MFE, which is efficient, dominant strategy incentive compatible and per-period individually rational, is unique.  We believe that these two properties might aid us in characterizing the equilibrium as time becomes large, and we defer this problem to future work.


\section{Value Determination}\label{sec:simulations}
\begin{figure*}[ht]
\centering
\begin{minipage}{.3\textwidth}
\centering
\includegraphics[width=1\columnwidth]{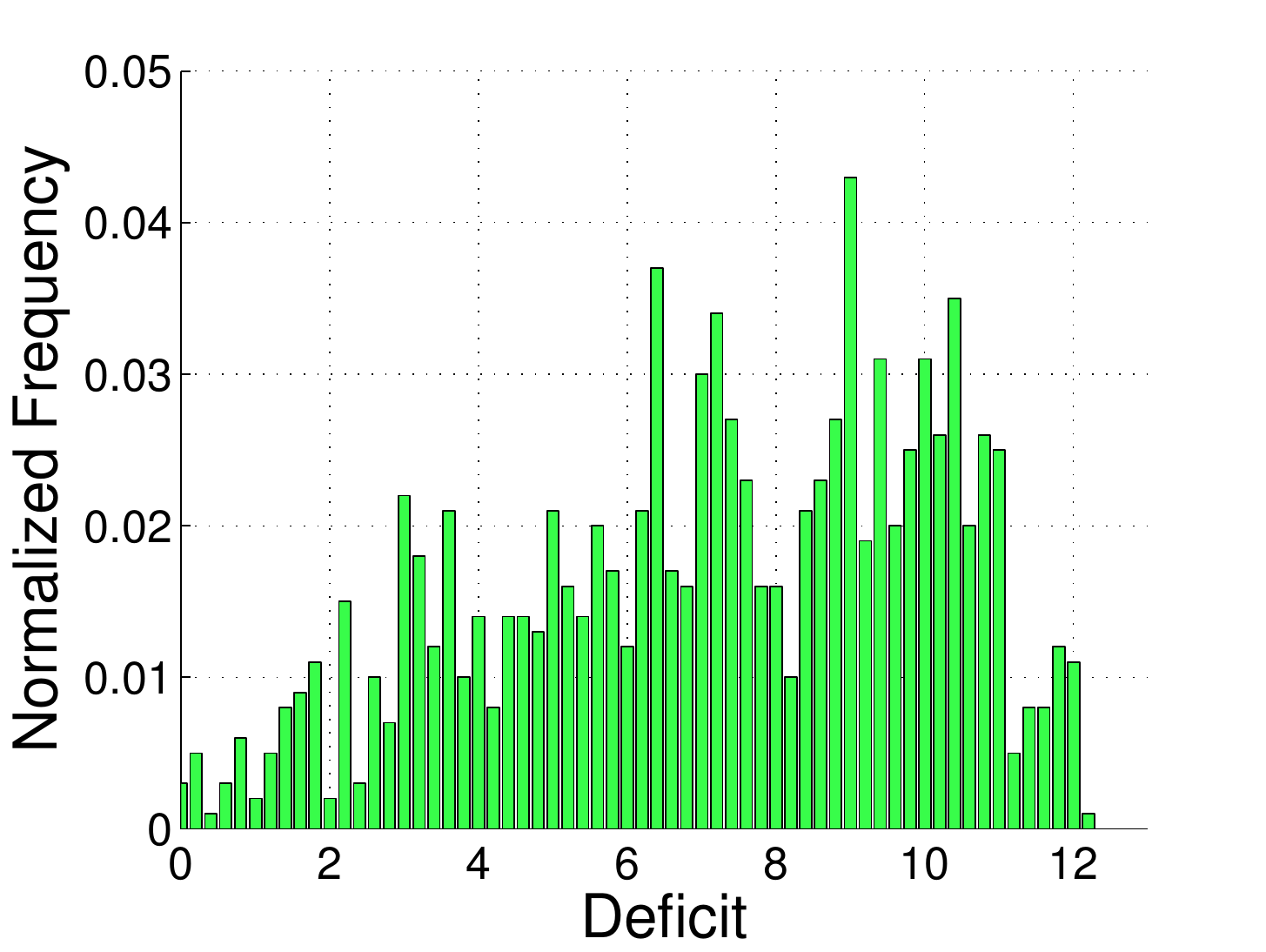}
\caption{Deficit distribution.}
\label{fig:rho}
\end{minipage}\hfill
\begin{minipage}{.3\textwidth}
\centering
\includegraphics[width=1\columnwidth]{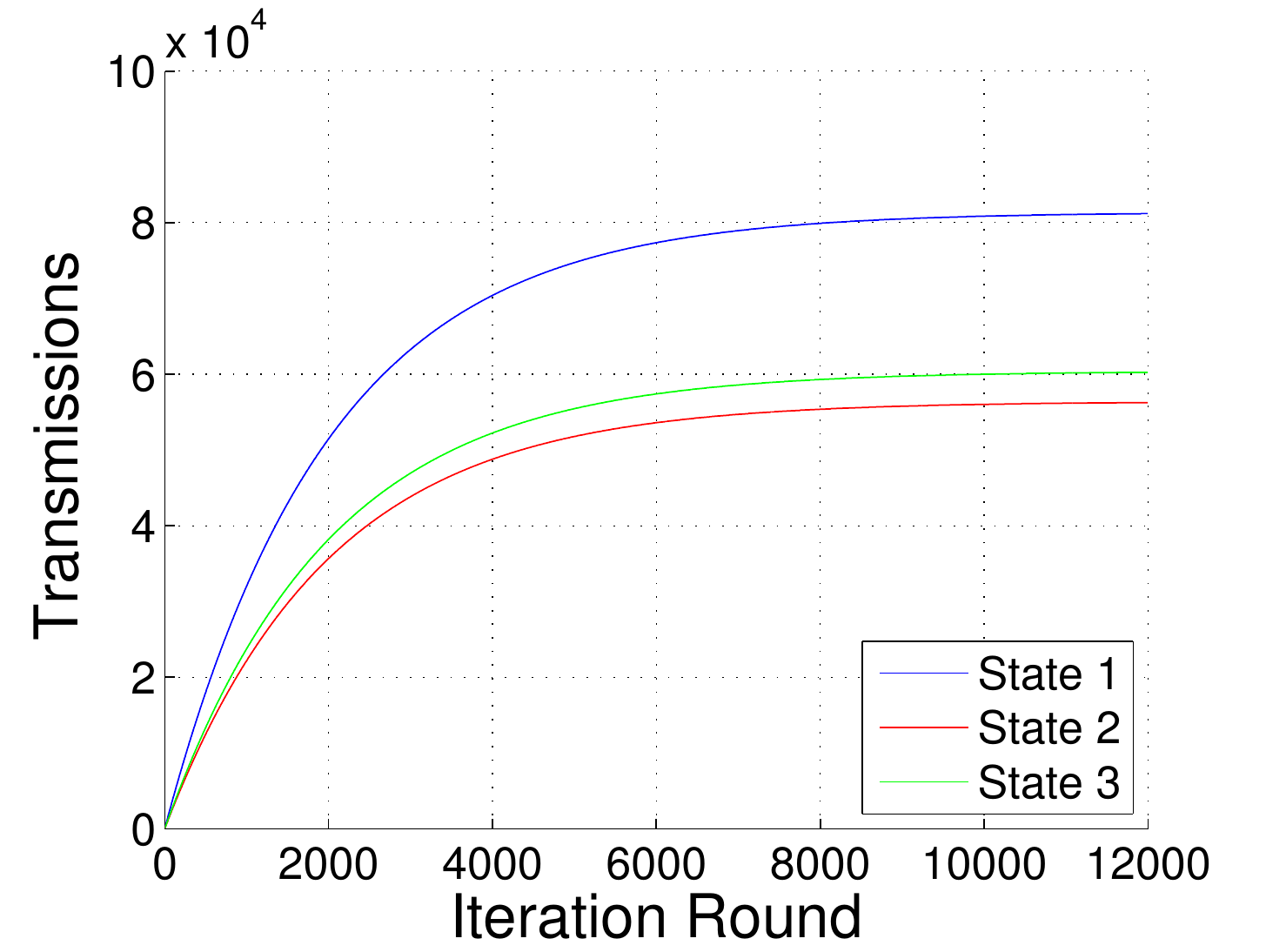}
\caption{Convergence of value iteration.}
\label{fig:conv}
\end{minipage}\hfill
\begin{minipage}{.3\textwidth}
\centering
\includegraphics[width=\columnwidth]{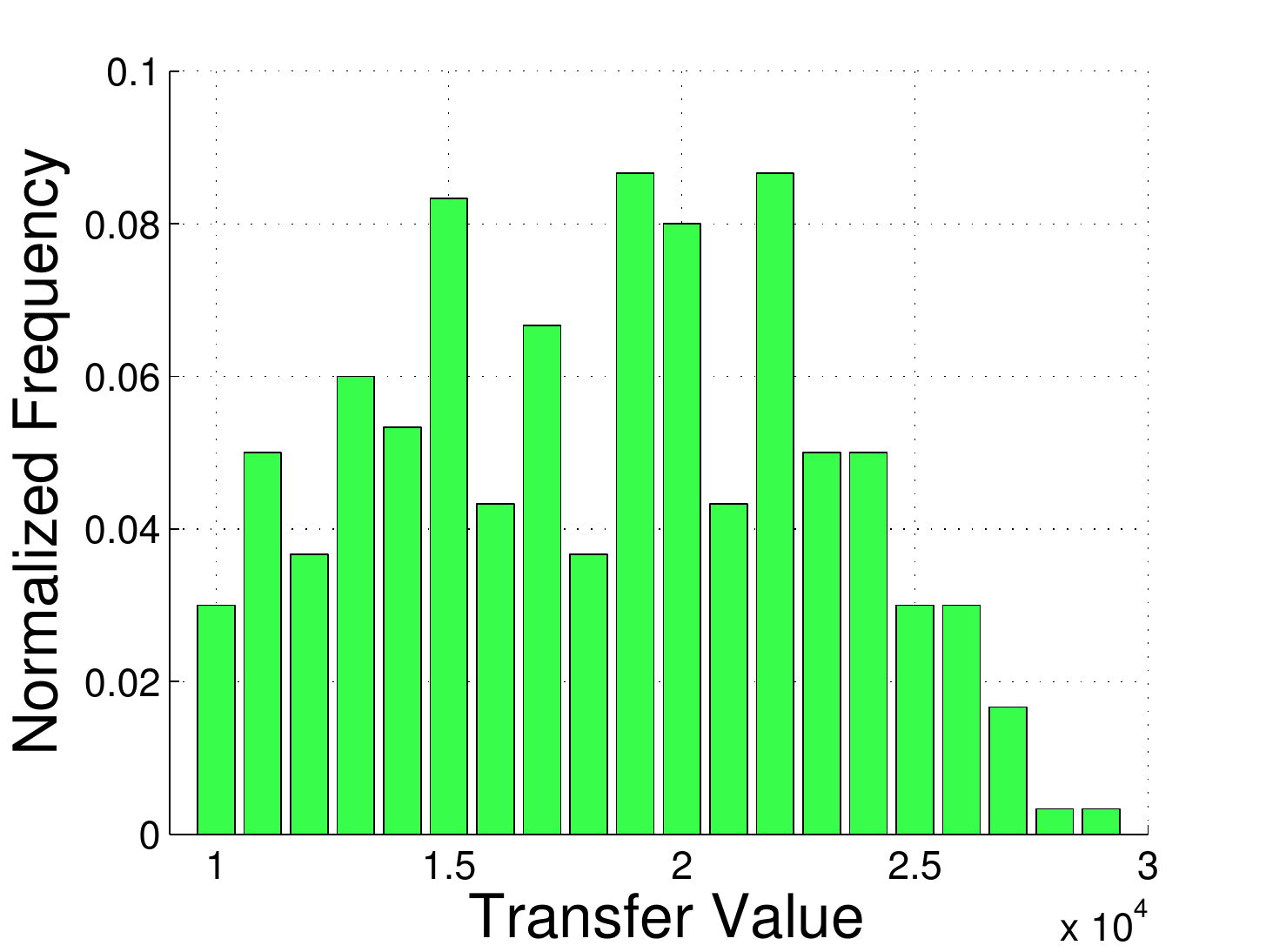}
\caption{Transfer distribution.}
\label{fig:transf}
\end{minipage}
\end{figure*}

We now turn to computing the system value from the viewpoint of a cluster and also a typical agent (say $1$). Here, we suppose there are $M=4$ agents in each cluster, and all have $\eta = 0.95,$ ${\delta} = 0.9995.$   Hence, each agent spends an average of $2000$ frames in the system before leaving.  A new agent has a deficit drawn uniformly at random from the interval $[0,13].$  Each agent needs to receive $N=10$ packets to decode the block, and there are $T=8$ time slots in each frame.  We wish to determine the value function from the perspective of the cluster and from the perspective of agent $1$, using \eqref{eq:valueMF} and \eqref{eq:valueAgentMF}.  

The following observation is useful to determine the allocations  $\mathbf{a}^*$ and $\tilde{\mathbf{a}}.$  It is straightforward to find $\mathbf{a}^*,$ since it simply follows Algorithm~\ref{alg: D2D}.  Now, consider  $\tilde{\mathbf{a}}.$   It is simple to see that it too would follow Phases $1$ and $2$ of Algorithm~\ref{alg: D2D}.   Then, from the perspective of agent $1,$ after the completion of these two phases, there are only two classes of allocations--those in which he transmits and those in which he does not.  Now, since all the other agents that agent $1$ comes in contact with in the future are drawn from $[\otimes \rho^{M-1}, \otimes \zeta^{M-1}],$ the allocation should follow a greedy minimization with respect to the other agents.  Thus, we only need consider two allocations while conducting value iterations: min-deficit-first with agent $1$ (identical to Phase $3$ of Algorithm~\ref{alg: D2D})  and min-deficit-first without agent $1$ (just set aside agent $1$ in Phase $3$ of Algorithm~\ref{alg: D2D}).

We first run the system according to  Algorithm~\ref{alg: D2D}, and use the results to find the empirical deficit distribution, denoted by $R.$  This is identical to the Mean Field deficit distribution.  The empirical distribution of deficit $R,$ is shown  in Figure \ref{fig:rho}.  We find that deficit lies in the range $0-13.$  

With $\eta=0.95$, the (countable) deficit set is $\{0, 0.05, 0.1, 0.15, ...\}.$  With a deficit range of $0-13$, there are totally $260$ potential values for deficit.  For the number of B2D chunks received $e$, we take values $3$, $4$ and $5$ (uniformly). Therefore, there are totally $260^4\times 3^4$ states in the system.  Using $R$ to represent the MF deficit distribution, and a linear holding cost function, we run value iteration; we present an example for a few states in Figure \ref{fig:conv}.   We thus obtain the mean field value functions. 

The empirical distribution of the average discounted transfers over the lifetime of each device is shown in Figure \ref{fig:transf}.  The average transfer is $18039.$  We will discuss the economic implications of this observation after describing the Android experiments in the next section.

\section{Android Implementation}\label{sec:android}
We now describe experiments on an Android testbed using a cluster size of four Google Nexus 7 tablets. We modified the kernel of Android v 4.3 to simultaneously allow both WiFi and 3G interfaces to transmit and receive data.  

Our system consists of a server application on a desktop that codes data and sends it to the tablets over the Internet,  an Android app that receives data over Internet on a 3G interface and shares it over the WiFi interface, and a monitor that keeps track of the state of the system and generates a trace of events.  The server initializes each tablet in the system with a randomly selected number of chunks.  Additionally, churn is emulated in the system by making the application on the tablet reset randomly with a probability $\bar{\delta} = 5 \times 10^{-4}$ (\textit{i.e.,} ${\delta}=0.9995$). 

We set the frame duration as $500$ ms.  Since we have ${\delta} = 0.9995,$ this means that the average duration that a device spends in the system is $1000$ seconds.   We use an MP3 music file as the data, and divide it into  blocks, with the blocks being further divided into chunks. Chunks are generated using an open source random linear coding library \cite{ncutils}, using field size $256$ and $10$ degrees of freedom per block.  Hence, a block is decodable with high probability if $10$ chunks are received successfully. Each chunk has an average size of $1500$ Bytes, and has a header that contains the frame number it corresponds to as well as its current deficit. The system maintains synchronization by observing these frame numbers.

The allocation algorithm proceeds as suggested by Algorithm~\ref{alg: D2D}.  We approximate the three phases by setting back-off times for D2D access.  Devices that cannot complete (\textit{i.e.}, Phase 1 devices) should be the most aggressive in D2D channel access.  We set them to randomly back-off between $1$ and $5$ ms before transmission. Devices that can afford to transmit some number of chunks (Phase 2) should be less aggressive, and transmit chunks by backing off between $1$ and $15$ ms.  Finally, each device enters Phase 3, and modulates its aggressiveness based on deficits. Each device normalizes its deficit based on the values of deficits that it sees from all other transmissions, and backsoff proportional to this deficit within the interval of $5$ to $15$ ms.  The average error in value due to a back-off based implementation is about $10-15\%$.

\begin{figure}[ht]
\begin {center}
\vspace{-0.1in}
\includegraphics[width=3.5in]{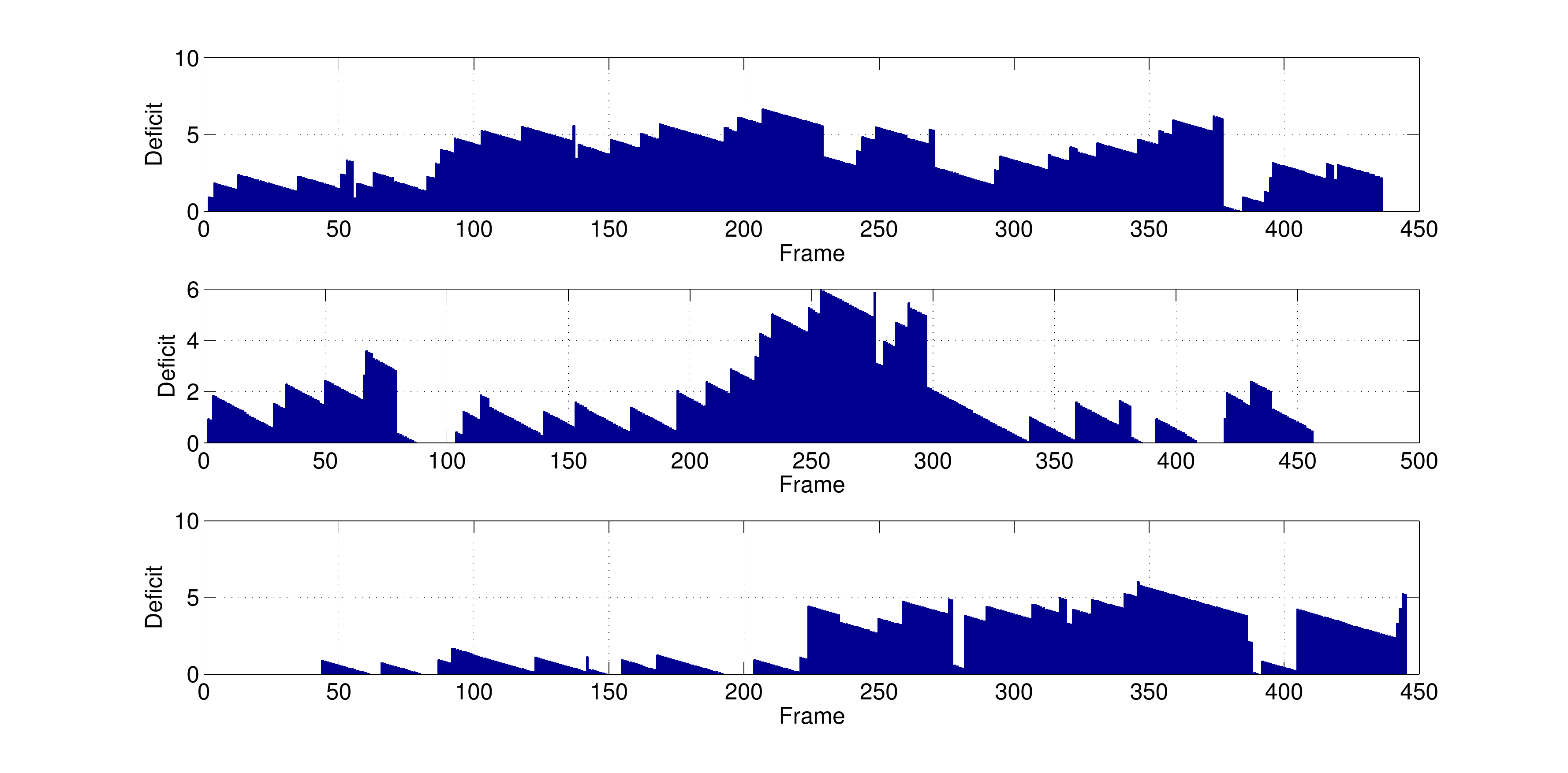}
\vspace{-0.3in}
\caption{Sample deficit trajectories.  We have used $\delta = 0.98$ in this run to illustrate frequent resets, which cause sharp decreases or increases.}
\label{fig:trajectory}
\vspace{-0.2in}
\end{center}
\end{figure}

We conducted experiments to determine the stable delivery ratio achieved using D2D for different B2D initializations per frame.  We present some sample deficit trajectories in Figure~\ref{fig:trajectory}.  The random resets emulating peer churn are visible as sharp changes in the deficit.   We found that on average, B2D transfer of $4$ chunks to each device is sufficient to ensure a delivery ratio of over $0.95.$  Hence, it is easy to achieve a $60\%$ reduction in B2D usage, while maintaining a high QoE.

\section{System Viability}\label{sec:viability}
We saw in Section~\ref{sec:simulations} that the average transfer to each agent is positive, meaning that the agents need to obtain some kind of subsidy in order to use the system.  What kind of subsidy should they be given?  The Android experiments indicate that each agent is able to save $60\%$ of the B2D costs when participating in the system.  Would this be sufficient?   

The price of B2D service is currently \$$10$ per GB across many US cellular providers.  Suppose that we consider music streaming at a rate of $250$ kbps corresponding to our Android system.  If each device uses only B2D communication (no D2D at all), the cost of spending $1000$ seconds in the sytem is $31.25$ cents.  The per frame communication cost is $0.0156$ cents, and we can consider this to be the value of each frame to the agent.

The experiments in Section~\ref{sec:android} indicate that the agents have to utilize their B2D connection for at least $40\%$ of the chunks to maintain the desired QoE.  Hence, the value that can potentially be received by participating in the D2D system is $0.6 \times 0.0156 = 0.00936$ cents per frame.   Let us assume a linear deficit cost function that takes a value of $0.00936$ cents at deficit value of $15.$  In other words, if the agent were to experience a deficit of $15$ or above in a frame, it gets no payoff from that frame.  Using this linear transformation, we can translate the average transfer of $18039$  (the value found in Section~\ref{sec:simulations}) over the entire $1000$ seconds into a total of $11.26$ cents.  Thus, if each agent saves at least $11.26$ cents, it has an incentive to participate in the D2D system.  The actual saving is $0.6*31.25 = 18.75$ cents ($60\%$ of the B2D costs) per agent, which is well above the minimum required saving.

The situation is still better for video streaming at a rate of $800$ kbps.  A similar calculation indicates that a $16$ minute video costs about $\$1$ using pure B2D,  while the B2D cost in the hybrid system is only $40$ cents, yielding a savings of $60$ cents per agent.  However, a saving of about $36$ cents per agent is all that is needed to incentivize them to participate.

In a full implementation, each agent would place an amount (\textit{eg.} $36$ cents for a $16$ minute average lifetime) in escrow with the monitor upon connecting.  Each agent would receive transfers according to our mechanism, and, on average, would receive its amount back from the monitor for its contributions.  Hence, the system would then be \textit{ex-ante} budget balanced.

\section{Conclusion}\label{sec:conclusion}
We studied the problem of providing incentives for cooperation in large scale multi-agent systems, using  wireless streaming networks as an example.  The objective was to incentivize truth telling about individual user states so that a system wide cost minimizing allocation can be used.  
We showed how a mean field approximation for large systems yields a low-complexity framework under which to design the mechanism.  
Finally, we implemented the system on Android devices and presented results illustrating its viability using the current price of cellular data access as the basis for transfers. 

\bibliographystyle{IEEEtran}
\bibliography{refs}  

\end{document}